\documentclass[numsec,webpdf,modern,medium,namedate]{oup-authoring-template}

\onecolumn

\usepackage{amsmath,amssymb,mathtools,bm}
\usepackage{graphicx}
\usepackage{booktabs,longtable}
\usepackage{siunitx}
\usepackage{placeins}       % \FloatBarrier
\usepackage{algorithm}
\usepackage{algpseudocode}
\usepackage{comment}
\usepackage{tikz}
\usetikzlibrary{positioning,arrows.meta,calc,fit,shapes.multipart,matrix}
\usepackage{rotating}       % sidewaystable, sidewaysfigure
\usepackage{threeparttable} % tablenotes
\usepackage{ragged2e}       % \justifying
\usepackage[hidelinks]{hyperref}
\usepackage[capitalise,noabbrev]{cleveref}

\theoremstyle{thmstyleone}%

\theoremstyle{thmstyletwo}%

\theoremstyle{thmstylethree}%
\newif\ifshowalttext
\showalttextfalse % set to true only for internal accessibility review

\newcommand{\alttext}[1]{%
  \par\smallskip\noindent\footnotesize\textit{Alt text: #1}\par
}
\begin{document}
% -----------------------------

\journaltitle{Journal of the Royal Statistical Society: Series C (Applied Statistics)}
%\DOI{DOI HERE}
\copyrightyear{2026}
\pubyear{2026}
%\access{Advance Access Publication Date: Day Month Year}
\appnotes{Original article}

\firstpage{1}

\title[Deep count regression]{Bayesian Deep Count Regression and Anomaly Detection: Evidence from GDELT Event Panels}

\author[1,$\ast$]{Hsin-Hsiung Huang}
\author[1]{Yuh-Haur Chen}
\author[1]{Mahlon Scott}

\address[1]{\orgdiv{School of Data, Mathematical, and Statistical Sciences},
\orgname{University of Central Florida},
\orgaddress{\street{4000 Central Florida Blvd}, \postcode{32816}, \state{FL}, \country{USA}}}

\corresp[$\ast$]{Address for correspondence. Hsin-Hsiung Huang, University of Central Florida, Orlando, FL 32816, USA. \href{mailto:hsin.huang@ucf.edu}{hsin.huang@ucf.edu}}

\received{28}{2}{2026}
%\revised{Date}{0}{Year}
%\accepted{Date}{0}{Year}

\abstract{
The Global Database of Events, Language and Tone (GDELT) provides geolocated event records that can be aggregated into weekly spatiotemporal panels of event counts across regions, actors, and event types. These panels are typically sparse, bursty, and overdispersed, so calibrated probabilistic forecasting is essential for monitoring rare surges. We propose Bayesian count regression pipelines that pair deterministic deep temporal encoders with negative binomial (NB2) and zero-inflated negative binomial (ZINB2) likelihood heads. Posterior predictive simulation yields predictive quantiles and right-tail probabilities that support both forecasting and anomaly scoring. For interpretable spillover attribution, we also fit a Bayesian generalised linear model with high-dimensional lagged cross-series predictors and a two-step screen-and-refit procedure under a three-parameter beta-normal (TPBN) shrinkage prior. To connect spillovers to directional statistics, active cross-region effects are mapped to geodesic bearings on the World Geodetic System 1984 ellipsoid (WGS84) and summarised using weighted circular moments, rose diagrams, and bearing-field maps. Simulations with known spillovers and conflict-panel case studies show accurate right-tail behaviour and a practical workflow for detecting and interpreting geopolitical shocks.
}


\boxedtext{
{\renewcommand{\theenumi}{\alph{enumi}}%
 \renewcommand{\labelenumi}{(\theenumi)}%
\begin{enumerate}
\item Hybrid deep count regression: deep temporal encoders paired with NB2 and ZINB2 likelihood heads to deliver calibrated posterior predictive distributions.
\item Sparse cross-series inference: a Bayesian GLM with TPBN shrinkage and two-step screening isolates a small set of active spillover drivers in high-dimensional panels.
\item Directional summaries on the sphere: active spillovers are mapped to geodesic bearings and summarised using posterior preferred bearings, rose diagrams, and bearing-field maps.
\end{enumerate}}%
}

\maketitle
\justifying

% =======================================================
\section{Introduction}
\label{sec:intro}
% =======================================================

Event databases such as the Global Database of Events, Language and Tone (GDELT) \citep{Leetaru13gdelt:global} provide structured, geolocated event records extracted from large-scale news sources. After aggregation, these records form large panels of nonnegative event count time series indexed by location, actor, event type, and time. Early warning and monitoring programs rely on such panels to forecast activity, quantify uncertainty, and flag statistically unusual event counts for analyst review.

Several features of these data complicate forecasting and monitoring. Many series contain long stretches of zeros punctuated by rare bursts, while others are dense but strongly overdispersed. Media coverage endogeneity, annotation changes, and language drift can introduce nonstationarities that resemble genuine changes in an event generating process. These properties challenge both generalized linear models (GLMs) and modern deep learning forecasters. Poisson or Gaussian residual assumptions often understate tail uncertainty, and purely data-driven sequence models may provide sharp point forecasts but poorly calibrated predictive distributions for rare event monitoring. Zero inflation is particularly important, motivating likelihoods such as zero-inflated Poisson \citep{lambertZeroInflatedPoissonRegression1992} or zero-inflated negative binomial \citep{minamiModelingSharkBycatch2007,he2024framework}.

A further operational complication is spatial spillover. Escalations in a focal region often co-occur with escalating activity in nearby regions and in global power centers. While spillover can be represented by a large set of cross-series predictors, interpretability requires geospatial attribution. Directional signals arise naturally when attributing inferred spillover effects to bearings on the sphere, from a focal region to the spatial origin of an estimated influence. This motivates directional statistical summaries and visual analytics on the sphere.

Motivated by the special issue focus on directional statistics, our contributions are threefold. First, we develop hybrid Bayesian deep count regression pipelines for sparse, bursty spatiotemporal panels by coupling a Temporal Fusion Transformer (TFT) \citep{lim2021temporal} or a Time Series Mixer (TSMixer) \citep{ekambaramTSMixerLightweightMLPMixer2023} encoder with a Bayesian NB2 or ZINB2 likelihood head. The encoder learns nonlinear temporal structure and cross-series representations, while the Bayesian likelihood head yields posterior predictive distributions used for forecasting, tail calibration assessment, and anomaly scoring.

Second, we present an interpretable Bayesian NB2 or ZINB2 GLM with high-dimensional cross-series predictors and TPBN shrinkage priors, estimated via a two-step screening and refitting procedure \citep{wangTwostepMixedtypeMultivariate2025}. This model yields sparse spillover attribution by identifying active source series and quantifying their posterior effects.

Third, we make the directional component explicit by converting inferred spillovers into directional summaries on the sphere. We map active source regions to geodesic bearings relative to a target region and construct posterior preferred-bearing summaries, directional rose diagrams, and bearing-field maps. These summaries connect regression-scale effects to interpretable geographic directions, aligning the empirical analysis with the directional statistics theme.

The remainder of the paper is organised as follows. \Cref{sec:data} describes the GDELT aggregation used to construct the panels. \Cref{sec:model} introduces the likelihoods, the sparse GLM with shrinkage, the deep-encoder likelihood-head hybrids, and the directional spillover summaries. \Cref{sec:inference} describes training, posterior inference, and anomaly scoring. \Cref{sec:sim} presents a simulation study with known spillovers. \Cref{sec:gdelt} presents case studies and directional visualisations. \Cref{sec:disc} concludes.

% =======================================================
\section{Dataset: Aggregated GDELT panels}
\label{sec:data}
% =======================================================

\subsection{Dataset introduction}
GDELT \citep{Leetaru13gdelt:global} provides geolocated event records with timestamps, CAMEO event codes, and actor fields. We construct spatiotemporal count panels by aggregating (i) the CAMEO root event code and (ii) the Actor 1 country code, and we spatially locate events using the Action geolocation field (the location of the event occurrence).

We analyze two weekly panels that both begin on the week of 23 February 2015 and extend through December 2025. The two panels differ only in their actor focus and the train-validation-test split used for model development and evaluation. For the Israel-Palestine (ISR) panel, the training period ends on 18 January 2021. For the Russia-Ukraine (UKR) panel, the training period ends on 20 January 2020. In both panels, the subsequent 52 weeks are used for validation, and all remaining weeks through December 2025 are used as a test set. This design places the major episodes discussed in our case studies (for example, the 2022 invasion period for UKR and the October 2023 escalation for ISR) in the held-out test period.

\begin{figure}[htbp]
    \centering
    \includegraphics[width=0.75\linewidth]{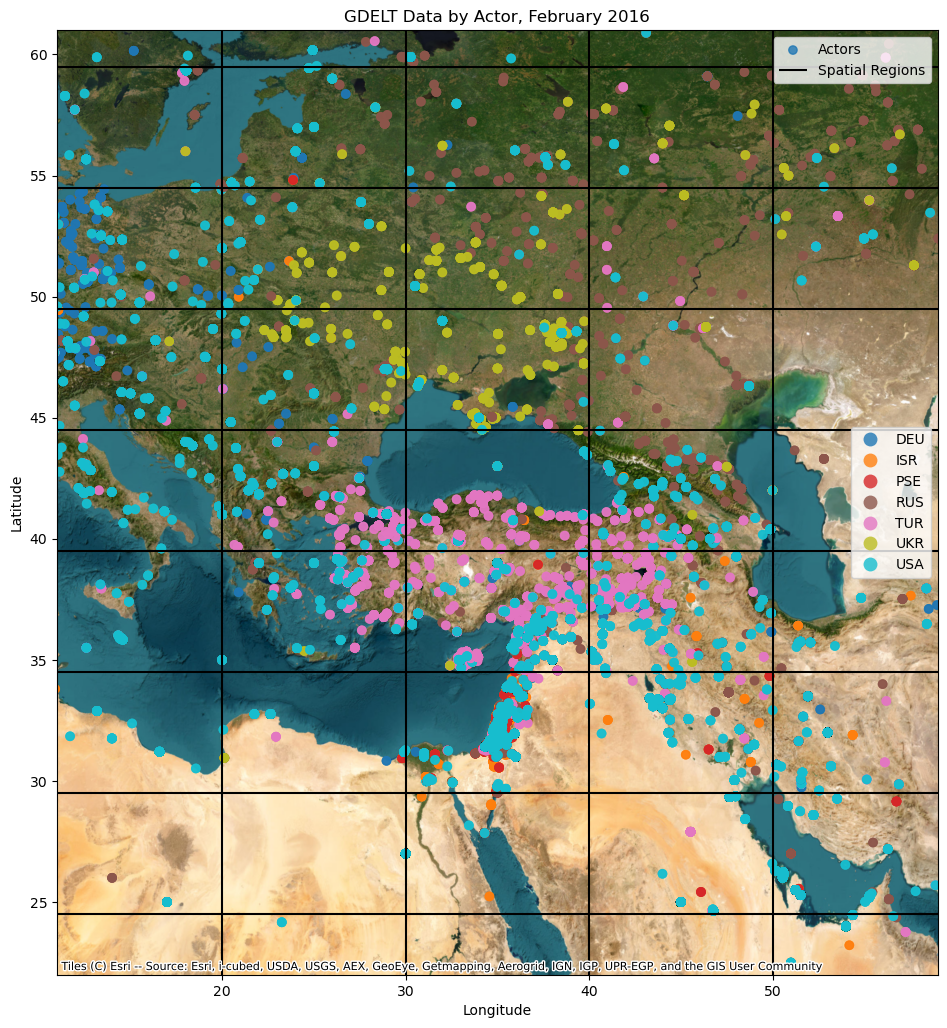}
    \caption{GDELT events by Action geolocation in February 2016, colored by Actor 1 country. Spatial regions are shown in black. Imagery source: Esri World Imagery.}
    \label{fig:gdeltdata}
    \smallskip
\alttext{Map of the study region showing a rectangular latitude-longitude grid and example event locations for one month, with points coloured by actor country.}
\end{figure}

Spatially, we restrict attention to a rectangular domain whose region centroids fall between 22 and 57 degrees latitude and between 15 and 55 degrees longitude. We discretize this domain into a regular grid of 5-degree latitude by 10-degree longitude cells, with centroids on the lattice
\[
\mathrm{lat}\in\{22,27,32,37,42,47,52,57\},\qquad
\mathrm{lon}\in\{15,25,35,45,55\},
\]
yielding $R=40$ spatial regions. Figure \ref{fig:gdeltdata} shows the resulting grid and an example month of raw event locations.

We restrict the actor dimension to the following set: United States, Russia, Ukraine, Israel, Palestine, Turkey, and Germany. Russia and Ukraine are included as the primary belligerents in the Russo-Ukrainian war, and Israel and Palestine are included for the Gaza war setting. Turkey and the United States are included as influential regional and global actors, and Germany is included as a proxy for broader European engagement. We additionally restrict the event dimension to four CAMEO root codes that span both cooperative and conflict-oriented actions: 04 (Consult), 13 (Threaten), 18 (Assault), and 19 (Fight) \citep{cameo}. The analysis in this paper uses only this four-code subset (rather than the full CAMEO root code set) to enable sufficiently long training runs for the Temporal Fusion Transformer (TFT) under the compute power available to us (2,000 H100 hours/month). This restriction, in turn, allows for a fairer comparison with TSMixer, which trains significantly faster and requires less GPU memory.

With $R=40$ regions, $A=7$ actors, and $V=4$ event codes, the panel contains $d = R\cdot A\cdot V = 1120$ parallel weekly count series.

\subsection{Event aggregation}
Let $r\in\{1,\dots,R\}$ index spatial regions, $t\in\{1,\dots,T\}$ index weekly time bins, $a\in\{1,\dots,A\}$ index Actor 1 countries, and $v\in\{1,\dots,V\}$ index CAMEO root event codes. Let $S_r$ denote the geographic cell for region $r$ and let $W_t$ denote the time interval for week $t$. For each combination $(r,t,v,a)$ we define the aggregated count
\begin{equation}
\label{eq:gdelt_aggregation}
Y_{r,t,v,a}
=
\sum_{i=1}^{N}
\mathbb{I}\Big\{
\mathrm{ActionGeo}(i)\in S_r,\;
\mathrm{Time}(i)\in W_t,\;
\mathrm{Actor1}(i)=a,\;
\mathrm{RootCode}(i)=v
\Big\},
\end{equation}
where $i$ indexes GDELT event records and $\mathbb{I}\{\cdot\}$ is the indicator function. Thus $Y_{r,t,v,a}\in\{0,1,2,\dots\}$ is the number of events of type $v$ attributed to Actor 1 country $a$ whose action location falls in region $r$ during week $t$.

\subsection{Spatial feature encoding}
To provide continuous spatial inputs for the deep encoders, we encode region locations using the WRAP representation \citep{9008116}. Let $c_r=(\mathrm{lat}_r,\mathrm{lon}_r)$ denote the centroid of region $r$ in degrees. The WRAP encoding is defined as
\[
f(c_r)=\left[
\sin\!\left(\pi\frac{\mathrm{lat}_r}{90}\right),\;
\cos\!\left(\pi\frac{\mathrm{lat}_r}{90}\right),\;
\sin\!\left(\pi\frac{\mathrm{lon}_r}{180}\right),\;
\cos\!\left(\pi\frac{\mathrm{lon}_r}{180}\right)
\right]^{\top}.
\]
This mapping ensures a simple, globally continuous encoding while retaining meaningful spatial information that allows the deep encoders to learn distances and similarities between regions.

% =======================================================
\section{Models}
\label{sec:model}
% =======================================================

Let $i=(r,v,a)$ index a target weekly count series and write $Y_{i,t}\equiv Y_{r,t,v,a}$ for week $t$. Our goal is to produce calibrated predictive distributions for $Y_{i,t+h}$ and to summarize spatial spillovers in a way that is interpretable on the sphere.

All pipelines in this paper share the same probabilistic likelihood (NB2 or ZINB2) and differ only in how the linear predictor is constructed. We consider three model classes:
(i) a sparse Bayesian GLM with high-dimensional cross-series predictors and TPBN shrinkage, fitted via a Two-Step screen-and-refit procedure \citep{wangTwostepMixedtypeMultivariate2025};
(ii) a TFT + GLM hybrid that uses a pre-trained Temporal Fusion Transformer as a nonlinear feature extractor and fits a Bayesian NB2/ZINB2 likelihood head;
(iii) a TSMixer + GLM hybrid that uses a pre-trained Time Series Mixer as a feature extractor with the same Bayesian likelihood head.

\subsection{Count likelihoods: NB2 and ZINB2}
\label{sec:bayes_glm}

We model $Y_{i,t}\in\{0,1,2,\dots\}$ using either a Negative Binomial type-2 (NB2) likelihood or a Zero-Inflated NB2 (ZINB2) likelihood. We select ZINB2 for a target series if at least 65\% of the responses in its training set are exactly zero; otherwise we use NB2. This rule is used only to choose the likelihood family and does not change the downstream anomaly scoring procedure.

\subsubsection{NB2 likelihood}
Let $\mu_{i,t}>0$ denote the conditional mean and let $\alpha_i>0$ denote the NB2 overdispersion parameter. Define $\kappa_i=\alpha_i^{-1}$ as the concentration. The NB2 pmf is
\begin{equation}
\label{eq:nb2_pmf}
P(Y_{i,t}=y\mid \mu_{i,t},\kappa_i)=
\frac{\Gamma(y+\kappa_i)}{\Gamma(\kappa_i)\,y!}
\left(\frac{\kappa_i}{\kappa_i+\mu_{i,t}}\right)^{\kappa_i}
\left(\frac{\mu_{i,t}}{\kappa_i+\mu_{i,t}}\right)^{y},
\end{equation}
with variance
\begin{equation}
\label{eq:nb2_var}
\mathrm{Var}(Y_{i,t}\mid \mu_{i,t},\alpha_i)=\mu_{i,t}+\alpha_i\mu_{i,t}^2.
\end{equation}
We link $\mu_{i,t}$ to covariates through a log link,
\begin{equation}
\label{eq:nb2_link}
\log \mu_{i,t}=\eta_{i,t}.
\end{equation}

\subsubsection{ZINB2 likelihood}
For sparse series we use a structural-zero gate $\pi_{i,t}\in(0,1)$:
\begin{equation}
\label{eq:zinb_model}
Y_{i,t} \sim
\begin{cases}
0, & \text{with probability }\pi_{i,t},\\
\mathrm{NB2}(\mu_{i,t},\alpha_i), & \text{with probability }1-\pi_{i,t}.
\end{cases}
\end{equation}
We model the gate using a logit link, $\mathrm{logit}(\pi_{i,t})=\eta^{(0)}_{i,t}$.
Implementation details for stable ZINB2 log-likelihood evaluation and posterior predictive simulation are provided in Appendix \ref{app:posterior_inference}.

\subsection{Sparse cross-series Bayesian GLM with TPBN shrinkage}
\label{sec:sparse_glm}

For each target series $i$ we construct (i) a low-dimensional block of unpenalized autoregressive predictors and (ii) a high-dimensional block of cross-series predictors. Let
\[
X_{i,t}=\Big[1,\ \log(1+Y_{i,t-1}),\ \log(1+Y_{i,t-2})\Big]^{\top}\in\mathbb{R}^{p},\qquad p=3,
\]
and let $Z_{i,t}\in\mathbb{R}^{q}$ collect lag-1 cross-series features of the form $\log(1+Y_{j,t-1})$ for other series $j\neq i$ in the panel (and any additional cross-series features included in a given experiment). In our panels $q$ can be large relative to the training length, motivating shrinkage.

\subsubsection{Linear predictors}
For NB2, the log-mean predictor is
\begin{equation}
\label{eq:sparse_glm_eta_nb}
\eta_{i,t}=X_{i,t}^{\top}\beta_i + Z_{i,t}^{\top}\gamma_i,
\end{equation}
where $\beta_i\in\mathbb{R}^{p}$ are unpenalized coefficients and $\gamma_i\in\mathbb{R}^{q}$ are shrinkage coefficients.
For ZINB2, we use separate predictors for the mean and the gate:
\begin{equation}
\label{eq:sparse_glm_eta_zinb}
\eta_{i,t}=X_{i,t}^{\top}\beta_i + Z_{i,t}^{\top}\gamma_i,
\qquad
\eta^{(0)}_{i,t}=X_{i,t}^{\top}\beta'_i + Z_{i,t}^{\top}\gamma'_i.
\end{equation}

\subsubsection{Priors}
We assign weakly informative Gaussian priors to unpenalized effects and a Gamma prior to the dispersion:
\[
\beta_i,\beta'_i \sim \mathcal{N}(0,\sigma^2 I),\qquad
\alpha_i \sim \mathrm{Gamma}(a_\alpha,b_\alpha).
\]
To induce sparsity in the high-dimensional cross-series effects, we place a three-parameter beta-normal (TPBN) shrinkage prior on each element of $\gamma_i$ (and $\gamma'_i$ in ZINB2) \citep{armagan2011generalized}.
\begin{align}
\gamma_{i,j} \mid \nu_{i,j} &\sim \mathcal{N}(0,\nu_{i,j}), \\
\nu_{i,j} \mid \eta_{i,j} &\sim \mathrm{Gamma}(u,\eta_{i,j}), \\
\eta_{i,j} \mid \tau_i &\sim \mathrm{Gamma}(a,\tau_i), \\
\tau_i &\sim \mathrm{HalfCauchy}(0,\tau_0),
\end{align}
with an independent copy of the same hierarchy for $\gamma'_{i,j}$ when fitting ZINB2. Posterior inference is performed with the No-U-Turn Sampler (NUTS) \citep{NUTS}.

\subsubsection{Two-Step screen-and-refit}
Following \citet{wangTwostepMixedtypeMultivariate2025}, we implement a Two-Step procedure to improve interpretability and reduce the shrinkage bias on large signals.
In Step 1, we fit the full model under TPBN shrinkage and select an active set of cross-series predictors using 95\% posterior credible intervals. Because TPBN is a continuous shrinkage prior, exact zeros do not occur in the posterior. To effectively filter out weak noise variables, we introduce a selection tolerance margin $\delta \ge 0$. Let $[L_{i,j}, U_{i,j}]$ denote the 95\% posterior credible interval for $\gamma_{i,j}$. For NB2, we define the active set by excluding variables whose credible intervals overlap with the equivalence region $[-\delta, \delta]$:
\[
\mathcal{A}_i=\{j:\ L_{i,j} > \delta \ \text{or}\ U_{i,j} < -\delta \}.
\]
For ZINB2, we use the union of the selected predictors across the mean and the zero-inflation gate:
\[
\mathcal{A}_i=\{j:\ (L_{i,j} > \delta \ \text{or}\ U_{i,j} < -\delta) \ \text{or}\ (L'_{i,j} > \delta \ \text{or}\ U'_{i,j} < -\delta) \}.
\]
In Step 2, we refit the same likelihood using only the reduced predictor set $\{Z_{i,t,j}:j\in\mathcal{A}_i\}$ with Gaussian priors (no TPBN) to obtain a sparse final model for forecasting, anomaly scoring, and directional summaries.

% =======================================================
\subsection{Directional spillover summaries on the sphere}
\label{sec:directional_glm}
% =======================================================

Directional information enters through geography: each cross-series predictor corresponds to a source region whose centroid lies on the Earth. For a target series $i$ in region $r_i$ and an active source predictor $j\in\mathcal{A}_i$ associated with source region $r_j$, we compute the forward azimuth (bearing) and geodesic distance from the source centroid to the target centroid,
\[
\omega_{j\rightarrow i}\in[0,2\pi),\qquad d_{j\rightarrow i}>0,
\]
using Vincenty's formulae \citep{Vincenty01041975} on the WGS84 ellipsoid \citep{wgs84}. Under this convention, $\omega_{j\rightarrow i}$ is the direction of the geodesic arrow drawn from the source toward the target. The direction from the target toward the source is the corresponding back-azimuth $(\omega_{j\rightarrow i}+\pi)\bmod 2\pi$.

We summarize posterior directional spillover using weighted circular moments. For posterior draw $s$, define weights
\[
w_{i,j}^{(s)}=\left|\gamma_{i,j}^{(s)}\right|
\quad\text{(or signed weights when producing sign-coded maps)}.
\]
The draw-specific resultant vector components are
\[
C_i^{(s)}=\sum_{j\in\mathcal{A}_i} w_{i,j}^{(s)}\cos(\omega_{j\rightarrow i}),
\qquad
S_i^{(s)}=\sum_{j\in\mathcal{A}_i} w_{i,j}^{(s)}\sin(\omega_{j\rightarrow i}),
\]
yielding a posterior draw of the preferred bearing
\begin{equation}
\label{eq:preferred_bearing}
\Omega_i^{(s)}=\mathrm{atan2}\!\left(S_i^{(s)},C_i^{(s)}\right),
\end{equation}
and a draw-specific directional concentration (mean resultant length)
\begin{equation}
\label{eq:mean_resultant_length}
R_i^{(s)}=
\frac{\sqrt{\left(C_i^{(s)}\right)^2+\left(S_i^{(s)}\right)^2}}
{\sum_{j\in\mathcal{A}_i} w_{i,j}^{(s)}}\in[0,1].
\end{equation}
We summarize $\Omega_i^{(s)}$ and $R_i^{(s)}$ across posterior draws to obtain preferred-bearing estimates and uncertainty bands, and we visualize the underlying bearing-weight pairs using (i) bearing-field maps that draw geodesic arrows from active source regions to the target and (ii) rose diagrams that bin bearings and aggregate weights within bins (\Cref{fig:spatial_map,fig:spatial_windrose}).

\subsection{Deep temporal encoders with Bayesian likelihood heads}
\label{sec:hybrid_model}

The sparse GLM is interpretable but remains linear in $X_{i,t}$ and $Z_{i,t}$. To benchmark against nonlinear representation learning while retaining calibrated predictive distributions, we use deep temporal encoders as feature extractors and fit the same Bayesian NB2/ZINB2 likelihood as a final layer.

\subsubsection{Two-stage hybrid construction}
Fix a lookback length $l$. For each target series $i$ and time $t$, let $x_{i,t}$ denote the multivariate historical input over the window $\{t-l,\dots,t-1\}$, together with any static identifiers used by the encoder (for example, region, actor, and event-code embeddings). A pre-trained encoder $g(\cdot)$ maps this history to an embedding
\[
h_{i,t}=g(x_{i,t})\in\mathbb{R}^{m},
\]
where $m$ is the embedding dimension (for TFT we use $m=24$; for TSMixer we use an embedding derived from the lookback dimension).
We then form a Bayesian likelihood head using unpenalized autoregressive lags and the learned embedding:
\begin{equation}
\label{eq:deep_head_eta}
\eta_{i,t}=\Big[1,\ \log(1+Y_{i,t-1}),\ \log(1+Y_{i,t-2}),\ h_{i,t}^{\top}\Big]\beta^{(\mathrm{enc})}_i,
\end{equation}
with $\beta^{(\mathrm{enc})}_i\sim\mathcal{N}(0,\sigma^2 I)$ and $\alpha_i\sim\mathrm{Gamma}(a_\alpha,b_\alpha)$. For sparse series, an analogous ZINB2 gate predictor $\eta^{(0)}_{i,t}$ is fitted using the same inputs. Importantly, we do not apply TPBN shrinkage to $h_{i,t}$ because the encoder output is designed to be dense.

\subsubsection{TFT + GLM likelihood head}
The Temporal Fusion Transformer (TFT) \citep{lim2021temporal} is trained to predict $\log(1+Y_{i,t})$ using a many-to-one formulation with static identifiers and shared temporal covariates. After training, we freeze the TFT parameters and compute $h_{i,t}$ for each $(i,t)$, then fit the Bayesian NB2/ZINB2 likelihood head \eqref{eq:deep_head_eta} via NUTS. This yields posterior predictive distributions that can be compared directly to the sparse GLM because the likelihood family and quantile-based anomaly scoring are identical.

\subsubsection{TSMixer + GLM likelihood head}
TSMixer \citep{ekambaramTSMixerLightweightMLPMixer2023} is trained in a many-to-many multivariate forecasting setup, learning a shared representation across all $d$ series in the panel. After training, we freeze the mixer and extract a target-specific embedding $h_{i,t}$ from the mixer output for series $i$ at time $t$. We then fit the same Bayesian NB2/ZINB2 likelihood head \eqref{eq:deep_head_eta} via NUTS.

\begin{figure}[t]
\centering
\begin{tikzpicture}[
  box/.style={draw, rectangle, rounded corners=1pt, inner sep=6pt, align=center},
  arrow/.style={-Latex, thick}
]
\node[box] (gdelt) {GDELT aggregation\\$\rightarrow$ panel counts $\{Y_{i,t}\}$};
\node[box, right=16mm of gdelt] (lik) {Likelihood choice\\(NB2 vs ZINB2)};
\node[box, right=16mm of lik] (sparse) {Sparse GLM\\TPBN + Two-Step\\Directional summaries};
\node[box, below=10mm of lik] (deep) {Deep encoder + GLM head\\(TFT or TSMixer)\\Bayesian NB2/ZINB2 head};
\node[box, below=10mm of sparse] (eval) {Posterior predictive\\forecasting and anomaly scoring};

\draw[arrow] (gdelt) -- (lik);
\draw[arrow] (lik) -- (sparse);
\draw[arrow] (lik) -- (deep);
\draw[arrow] (sparse) -- (eval);
\draw[arrow] (deep) -- (eval);
\end{tikzpicture}
\caption{Modeling workflow. All pipelines share the same count likelihood (NB2 or ZINB2). The sparse GLM uses TPBN shrinkage and a Two-Step refit for interpretable cross-series spillovers, which are mapped to geodesic bearings for directional summaries. The hybrid pipelines replace high-dimensional cross-series predictors with learned embeddings from TFT or TSMixer and fit a Bayesian likelihood head for calibrated posterior predictive distributions.}
\label{fig:pipeline_dir_bayes}
\smallskip
\alttext{Flow diagram of the modelling workflow from panel aggregation to likelihood choice, then either sparse regression with directional summaries or deep encoder with likelihood head, ending in forecasting and anomaly scoring.}
\end{figure}
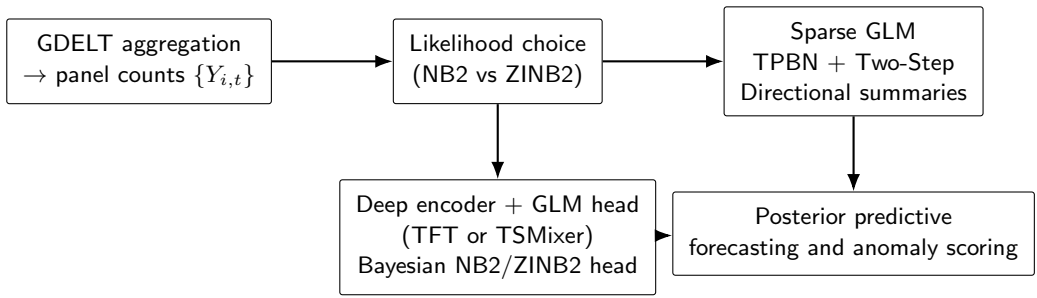

\begin{algorithm}[tbp]
\caption{TFT + GLM hybrid Bayesian inference pipeline}
\label{alg:tft_hybrid}
\begin{algorithmic}[1]
\Require Target series $\{Y_{i,t}\}$, historical inputs $\{x_{i,t}\}$, pre-trained TFT encoder $g_{\mathrm{TFT}}$, choice of NB2 or ZINB2 head.
\Statex \textbf{Step 1: Embedding extraction}
\State Compute embeddings $h_{i,t}=g_{\mathrm{TFT}}(x_{i,t})$ for all required $(i,t)$.
\Statex \textbf{Step 2: Bayesian likelihood head}
\State Construct $\eta_{i,t}$ using \eqref{eq:deep_head_eta} and fit NB2 or ZINB2 via NUTS to obtain posterior samples.
\Statex \textbf{Step 3: Posterior predictive}
\State Simulate posterior predictive draws to obtain predictive quantiles and right-tail probabilities for anomaly scoring.
\end{algorithmic}
\end{algorithm}

% =======================================================
\section{Training, posterior inference, and anomaly scoring}
\label{sec:inference}
% =======================================================

This section describes how we fit the hybrid deep pipelines and the sparse Bayesian GLM, and how we convert posterior predictive draws into rolling forecasts and anomaly scores for panel monitoring.

\subsection{Two-stage estimation for the hybrid deep models}
The TFT + GLM and TSMixer + GLM pipelines are estimated in two stages.

\subsubsection{Stage 1: train a deterministic encoder}
Let $i=(r,v,a)$ index a target series and let $Y_{i,t}$ denote its weekly count at week $t$. For a lookback window length $l$, let $x_{i,t}$ denote the multivariate history used by the encoder at forecast origin $t$ (all required series over weeks $t-l,\dots,t-1$, together with any static identifiers). We train the encoder parameters $\theta$ by minimizing mean absolute error on a variance-stabilized scale,
\[
\min_{\theta}\ \frac{1}{|\mathcal{I}||\mathcal{T}_{\mathrm{train}}|}
\sum_{i\in\mathcal{I}}\sum_{t\in\mathcal{T}_{\mathrm{train}}}
\left|\log(1+Y_{i,t})-\widehat{y}_{i,t}(\theta)\right|,
\]
where $\widehat{y}_{i,t}(\theta)$ is the encoder's point prediction for $\log(1+Y_{i,t})$ and $\mathcal{I}$ is the set of series included in the panel.

\subsubsection{Stage 2: freeze the encoder and fit a Bayesian likelihood head}
After training, we freeze $\hat\theta$ and compute embeddings
\[
h_{i,t}=g_{\hat\theta}(x_{i,t}).
\]
We then fit a Bayesian NB2 or ZINB2 likelihood head using the linear predictor defined in \Cref{sec:hybrid_model}. The likelihood head is fit with NUTS, producing posterior samples of the head parameters and posterior predictive draws for forecasting and anomaly scoring.

\subsubsection{Reducing in-sample optimism for the likelihood head}
The hybrid pipelines treat $h_{i,t}$ as fixed in Stage 2. This ignores uncertainty in the encoder weights and can lead to optimistic uncertainty quantification if the likelihood head is fit using embeddings computed only on the same weeks used to train the encoder. To partially mitigate this effect, we use the train, validation, and test split described in \Cref{sec:data} as follows.

Let $\mathcal{T}_{\mathrm{train}}$ and $\mathcal{T}_{\mathrm{val}}$ denote the training and validation weeks, and let $\mathcal{T}_{\mathrm{test}}$ denote the held-out test weeks. We train the encoder using only $\mathcal{T}_{\mathrm{train}}$. We then fit the Bayesian likelihood head using embeddings computed for all weeks in $\mathcal{T}_{\mathrm{train}}\cup\mathcal{T}_{\mathrm{val}}$, where each $h_{i,t}$ is computed using only information available at time $t-1$ (rolling origin, no leakage). Including validation-period embeddings exposes the likelihood head to realistic out-of-sample representation error and improves the calibration of posterior predictive tails used for monitoring.

\subsection{Posterior inference via Hamiltonian Monte Carlo}
All Bayesian likelihood heads and sparse GLMs are fit using Hamiltonian Monte Carlo with the No-U-Turn Sampler (NUTS) \citep{NUTS}. We implement inference in \texttt{NumPyro} \citep{phan2019composable, bingham2019pyro}, which supports automatic differentiation and efficient parallel chains.

For the sparse GLM with TPBN shrinkage (\Cref{sec:sparse_glm}), we use the Two-Step screen-and-refit workflow \citep{wangTwostepMixedtypeMultivariate2025} detailed in Algorithm \ref{alg:twostep_nuts}. In Step 1, NUTS samples the full shrinkage model and yields posterior credible intervals for screening. In Step 2, NUTS refits a reduced model with standard Gaussian priors to obtain a sparse, interpretable posterior for forecasting and directional summaries. Numerical stabilization choices for NB2 and ZINB2 log densities, and the shrinkage reparameterizations used to improve sampling geometry, are provided in Appendix \ref{app:posterior_inference}.

In all Bayesian fits we monitor standard MCMC diagnostics (effective sample sizes and $\widehat{R}$ across chains) and retain posterior draws only after warmup and adaptation.

\subsection{Rolling forecasts and anomaly scoring}
\label{sec:anomaly_scoring}

For each target series $i$ and forecast origin $t$, we summarize uncertainty and anomalous behavior through posterior predictive simulation under the fitted NB2 or ZINB2 model. Let $\mathcal{F}_{t-1}$ denote the information available through week $t-1$, let $\mathcal{D}_{t-1}$ denote the data used to fit the model that produces the forecast at $t$, and let $\Theta_i$ collect the model parameters for series $i$. The one-step-ahead posterior predictive distribution is
\[
p(Y_{i,t}\mid \mathcal{F}_{t-1})
=\int p(Y_{i,t}\mid \Theta_i,\mathcal{F}_{t-1})\,p(\Theta_i\mid \mathcal{D}_{t-1})\,d\Theta_i.
\]
Given posterior draws $\{\Theta_i^{(s)}\}_{s=1}^{S}$, we approximate this distribution by simulating
$\tilde{Y}_{i,t}^{(s)} \sim p(\,\cdot\mid \Theta_i^{(s)},\mathcal{F}_{t-1})$,
yielding a Monte Carlo sample $\{\tilde{Y}_{i,t}^{(s)}\}_{s=1}^{S}$.

From these posterior predictive draws we compute (i) dynamic predictive bounds, (ii) binary anomaly flags, and (iii) continuous tail-probability scores. For a fixed right-tail level $q\in(0,1)$, define the predictive upper bound $U_{i,t,q}$ as the empirical $q$-quantile of $\{\tilde{Y}_{i,t}^{(s)}\}$. The corresponding right-tail flag is
\begin{equation}
\label{eq:anomaly_flag}
A_{i,t}(q)=\mathbb{I}\!\left\{Y_{i,t} > U_{i,t,q}\right\},
\end{equation}
and we use $q=0.975$ throughout. A continuous anomaly score is given by the posterior predictive right-tail probability
\begin{equation}
\label{eq:anomaly_score}
p_{i,t}=P(\tilde{Y}_{i,t}\ge Y_{i,t}\mid \mathcal{F}_{t-1})
\approx \frac{1}{S}\sum_{s=1}^{S}\mathbb{I}\!\left\{\tilde{Y}_{i,t}^{(s)}\ge Y_{i,t}\right\}.
\end{equation}
Smaller $p_{i,t}$ indicates a more extreme right-tail realization under the fitted model and provides a natural ranking for analyst triage. Thresholding $p_{i,t}$ is equivalent to the quantile rule in \eqref{eq:anomaly_flag}.

For the sparse GLM, flagged times can be interpreted geographically by examining the posterior of active cross-series coefficients and mapping them to geodesic bearings (\Cref{sec:directional_glm}). We summarize the implied bearing-weight field using posterior preferred-bearing summaries and rose diagrams, and we optionally report the largest contributing active sources for the flagged window.

To quantify calibration of the right tail on held-out data, we report the exceedance-rate deviation
\begin{equation}
\label{eq:tail_calibration_metric}
T(q)
=
\left|
(1-q)
-
\frac{1}{N}\sum_{n=1}^{N}
\mathbb{I}\!\left\{Y_{i_n,t_n}>U_{i_n,t_n,q}\right\}
\right|,
\end{equation}
computed over $N$ held-out series-week pairs $\{(i_n,t_n)\}_{n=1}^{N}$. Smaller $T(q)$ indicates that the empirical exceedance frequency matches the nominal level $1-q$ more closely.

\subsection{Computational remarks}
The three model families differ substantially in computational scaling because they treat the panel structure differently.

We represent the panel as $Y\in\mathbb{R}^{T\times d}$ with $d=R\cdot A\cdot V$ series and use historical input tensors $X\in\mathbb{R}^{T\times l\times d}$ for multivariate encoders. TFT training uses a flattened many-to-one format, which replicates shared temporal covariates across series identifiers and leads to higher memory and compute cost as $d$ increases. TSMixer uses a many-to-many multivariate format that avoids this replication and is therefore substantially cheaper to train in our setting.

The sparse Bayesian GLM fits a separate probabilistic model for each series (with parallelization across series), while NUTS sampling cost depends on posterior geometry and the number of parameters. The Two-Step procedure reduces the dimensionality of the final refit, improving interpretability and often improving sampling efficiency in the second-stage model.

\section{Numerical studies: simulation}
\label{sec:sim}
% =======================================================

\subsection{Simulation design and data generation}
We evaluate the proposed Two-Step Shrinkage GLM against representation-learning baselines and standard GLM benchmarks using a controlled simulation that mimics three features that recur in GDELT-style panels: short-range temporal dependence, cross-series spillovers with a sparse set of true drivers, and strong overdispersion with occasional structural zeros.

We generate $T=1000$ weekly observations and split them into $T_{\mathrm{train}}=950$ training weeks and $T_{\mathrm{test}}=50$ held-out test weeks. The system contains $100$ parallel series: one dense target series, one sparse target series, and $98$ additional exogenous noise series. The $98$ noise series are drawn independently as
\[
U_{j,t}\sim \mathrm{Poisson}(1.5), \qquad j=1,\dots,98,\ \ t=1,\dots,T,
\]
and all lagged inputs entering linear predictors are transformed with $\log(1+x)$ to reduce the influence of extreme values.

The dense target $Y^{(D)}_t$ is generated from an NB2 model. Conditional on its mean $\mu^{(D)}_t$,
\[
Y^{(D)}_t \sim \mathrm{NB2}(\mu^{(D)}_t,\alpha_D),
\]
where NB2 is the mean-dispersion parameterization used in \Cref{sec:bayes_glm}. The mean follows an AR(2) structure plus a high-dimensional lag-1 cross-series component:
\begin{equation}
\label{eq:sim_dense_mean}
\log \mu^{(D)}_t
=
\beta_{0,D}
+
\beta_{1,D}\log(1+Y^{(D)}_{t-1})
+
\beta_{2,D}\log(1+Y^{(D)}_{t-2})
+
\sum_{k\in\mathcal{A}}
\gamma_{k,D}\, \log(1+W_{k,t-1}),
\end{equation}
where $W_{k,t-1}$ ranges over the lag-1 value of the sparse target and the $98$ noise series, yielding $99$ exogenous candidates in total. The active set $\mathcal{A}$ has exactly $|\mathcal{A}|=5$ elements, so the true cross-series dependence is sparse.

The sparse target $Y^{(S)}_t$ is generated from a ZINB2 model with both a mean component and a structural-zero gate:
\[
Y^{(S)}_t \sim
\begin{cases}
0, & \text{with probability }\pi_t,\\
\mathrm{NB2}(\mu^{(S)}_t,\alpha_S), & \text{with probability }1-\pi_t.
\end{cases}
\]
We use parallel regressions for $\mu^{(S)}_t$ and $\pi_t$:
\begin{align}
\label{eq:sim_sparse_mean}
\log \mu^{(S)}_t
&=
\beta_{0,S}
+
\beta_{1,S}\log(1+Y^{(S)}_{t-1})
+
\beta_{2,S}\log(1+Y^{(S)}_{t-2})
+
\sum_{k\in\mathcal{A}}
\gamma_{k,S}\,\log(1+V_{k,t-1}),
\\
\label{eq:sim_sparse_gate}
\mathrm{logit}(\pi_t)
&=
\beta'_{0,S}
+
\beta'_{1,S}\log(1+Y^{(S)}_{t-1})
+
\beta'_{2,S}\log(1+Y^{(S)}_{t-2})
+
\sum_{k\in\mathcal{A}}
\gamma'_{k,S}\,\log(1+V_{k,t-1}),
\end{align}
where $V_{k,t-1}$ ranges over the lag-1 value of the dense target and the $98$ noise series (again $99$ candidates total). The same active index set $\mathcal{A}$ is used in both \eqref{eq:sim_sparse_mean} and \eqref{eq:sim_sparse_gate}, so the sparse target has a coherent set of true drivers across its count and zero-inflation components.

To avoid numerical overflow in generation and downstream fitting, we clip linear predictors prior to exponentiation when producing $\mu^{(D)}_t$ and $\mu^{(S)}_t$. This mirrors the stabilization choices used in the empirical pipelines (Appendix \ref{app:posterior_inference}).

\subsection{Model pipelines compared}
\label{sec:sim_pipelines}

We compare seven pipelines for each target type (NB2 for the dense target $Y^{(D)}_t$ and ZINB2 for the sparse target $Y^{(S)}_t$). All Bayesian likelihood-based models are fit with the No-U-Turn Sampler (NUTS) using multiple chains and a warmup period. For the deep-learning pipelines, estimation is two-stage: (i) the encoder is trained on the training split, and (ii) a Bayesian NB2 or ZINB2 likelihood head is fit on top of the frozen encoder outputs so that uncertainty and tail probabilities are produced by the same parametric count likelihood across pipelines.

\begin{enumerate}
\item \textbf{True DGP (oracle).}
This benchmark does not estimate parameters. Instead, it evaluates the predictive distribution implied by the exact data-generating parameters used in \Cref{sec:sim} (including autoregressive terms, active cross-series effects, and dispersion). The oracle quantifies irreducible uncertainty and provides a reference for both interval width and finite-sample exceedance variability.

\item \textbf{AR2 baseline GLM.}
A minimal Bayesian benchmark with only an intercept and two autoregressive lags of the target series. This pipeline isolates what can be achieved without cross-series information.

\item \textbf{Full GLM.}
A high-dimensional Bayesian GLM that uses the AR2 block plus all $99$ exogenous candidates as fixed effects, with weakly informative Gaussian priors and no shrinkage or screening. This pipeline illustrates the instability that can arise when many weak predictors are included without structured regularization.

\item \textbf{TFT-Bayesian head.}
A representation-learning pipeline in which a Temporal Fusion Transformer (TFT) is trained as a multivariate encoder and then used to produce a dense embedding for the target at each forecast origin. We then fit a Bayesian NB2 or ZINB2 likelihood head using the AR2 block plus the TFT embedding (dimension $24$ in our implementation). The shrinkage block is empty in this pipeline because the embedding coordinates are dense learned features rather than interpretable candidate drivers.

\item \textbf{TSMixer-Bayesian head.}
A second representation-learning pipeline in which a Time Series Mixer (TSMixer) encoder is trained and then frozen. The Bayesian likelihood head uses the AR2 block plus the target-specific TSMixer embedding (dimension determined by the lookback window). As in the TFT pipeline, the shrinkage block is empty so that calibration reflects the encoder representation rather than additional screening.

\item \textbf{Two-Step Shrinkage GLM.}
An interpretable high-dimensional regression pipeline. The fixed-effects block contains only the AR2 terms (and intercept), while the $99$ exogenous candidates enter through a shrinkage block under the TPBN prior. We apply the Two-Step screen-and-refit procedure to obtain a sparse final model and a posterior predictive distribution from the refitted likelihood head.

\item \textbf{TSMixer + Shrinkage.}
A hybrid wide-and-deep variant designed to test whether representation learning and sparse screening can be combined without loss of calibration. The AR2 block and the TSMixer embedding enter as fixed effects, while the raw $99$ exogenous candidates enter the TPBN shrinkage block and are screened via the same Two-Step procedure. This pipeline evaluates whether the additional screened covariates provide complementary information beyond the learned embedding, or whether the two components compete to explain overlapping variation.
\end{enumerate}

All seven pipelines are evaluated on the same held-out test window using the same forecast horizon and the same posterior predictive summaries (median forecasts, predictive intervals, and exceedance-based tail metrics). This design makes performance differences attributable to (i) sparse screening of candidate cross-series drivers versus (ii) learned representations from deep encoders, rather than to differences in the probabilistic output layer.

\subsection{Evaluation metrics}\label{sec:metrics}
All metrics are computed on the final $T_{\mathrm{test}}=50$ held-out weeks. Point forecast accuracy is summarized by mean absolute error on the raw count scale and on a stabilized log scale:
\[
\mathrm{MAE}_{\mathrm{raw}}=\frac{1}{50}\sum_{t\in\mathcal{T}_{\mathrm{test}}}\left|Y_t-\widehat{Y}_t\right|,
\qquad
\mathrm{MAE}_{\log}=\frac{1}{50}\sum_{t\in\mathcal{T}_{\mathrm{test}}}\left|\log_{10}(1+Y_t)-\log_{10}(1+\widehat{Y}_t)\right|,
\]
where $\widehat{Y}_t$ is the posterior predictive median for Bayesian models (and the model median implied by the oracle for True DGP).

For anomaly monitoring we focus on the predictive upper quantile at $q=0.975$. Let $U_{t,0.975}$ denote the posterior predictive 97.5\% quantile. We summarize right-tail calibration by
\begin{equation}
\label{eq:sim_tail_metric}
T
=
\left|
0.025 - \frac{1}{50}\sum_{t\in\mathcal{T}_{\mathrm{test}}}\mathbb{I}\{Y_t>U_{t,0.975}\}
\right|.
\end{equation}
Smaller $T$ indicates better matching between nominal and empirical exceedance rates. Because the test set has length $50$, the expected number of exceedances under perfect calibration is $50\times 0.025=1.25$, so the realized exceedance count must be an integer and exact equality to $0.025$ is generally not attainable in a single finite run.

\subsection{Numerical results}
\label{sec:sim_results}

The quantitative results for both targets are summarised in Table \ref{tab:sim_results}. We include the oracle (True DGP) to provide a finite-sample reference for both point accuracy and right-tail exceedances, and we report the remaining pipelines to highlight trade-offs between representation learning and sparse screening under high-dimensional noise.

Several patterns are consistent across targets.

{\renewcommand{\theenumi}{\alph{enumi}}%
 \renewcommand{\labelenumi}{(\theenumi)}%
\begin{enumerate}
\item \textit{Oracle tail behaviour provides a finite-sample reference.}
Even under the true data-generating mechanism, exceedances are random and integer-valued over a 50-step window. In this run, the dense target exhibits three exceedances (so $T=|0.025-3/50|=0.035$) and the sparse target exhibits zero exceedances (so $T=0.025$). These oracle values provide a realistic benchmark for interpreting the fitted models.

\item \textit{Two-Step Shrinkage reproduces the oracle right tail while remaining competitive in point accuracy.}
For both targets, the Two-Step Shrinkage GLM matches the oracle $T$ values in this run, indicating that the predictive upper tail is not systematically too liberal or too conservative relative to the mechanism used to generate the data. Point accuracy remains close to the oracle on the dense target and is competitive on the sparse target.

\item \textit{Unregularised high-dimensional regression inflates uncertainty and degrades accuracy in the sparse regime.}
The Full GLM performs poorly for the sparse target in both MAE metrics. Its favourable $T$ value in this run is consistent with overly wide predictive bands that suppress exceedances but provide weak operational thresholds.

\item \textit{Representation learning can improve point forecasts, but tail behaviour depends on the learned embedding and its interaction with the count likelihood.}
The TFT-based pipeline achieves strong dense-target point forecasts in this run and exhibits tail behaviour comparable to the oracle. The TSMixer-based likelihood head is less well calibrated in the right tail for both targets, yielding more exceedances than expected.

\item \textit{Stacking embeddings with a shrinkage block can be unstable when the two components explain overlapping variation.}
The TSMixer plus shrinkage variant improves sparse-target tail behaviour relative to TSMixer alone, but it remains worse than Two-Step Shrinkage and degrades dense-target point accuracy, consistent with collinearity between expressive embeddings and raw exogenous candidates.
\end{enumerate}}

\subsection{Dynamic uncertainty and interval sharpness}
\label{sec:sim_sharpness}

\begin{figure}[htbp]
\centering
\includegraphics[width=0.9\textwidth]{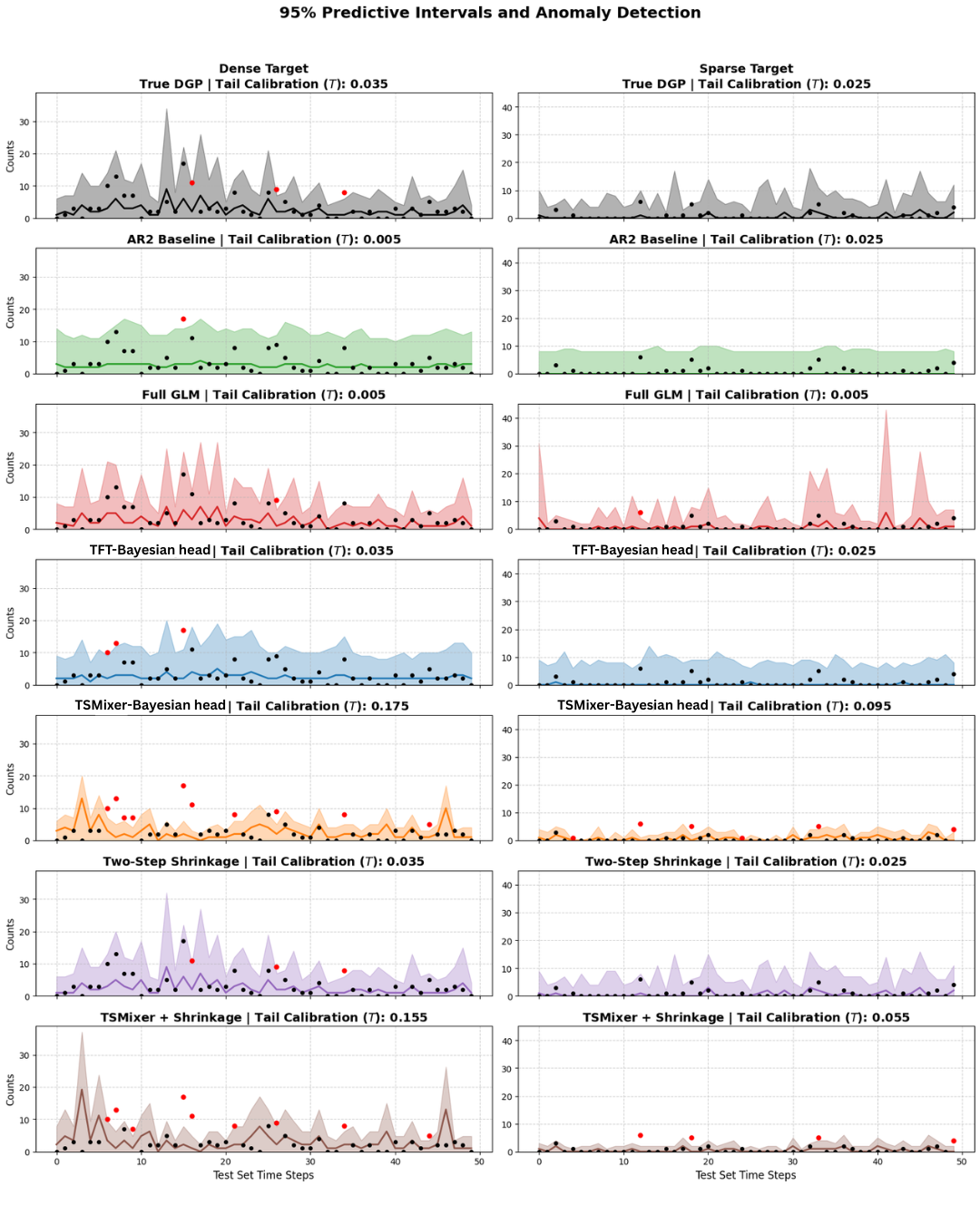}
\caption{Out-of-sample predictive performance on the 50-step test set for Dense (left column) and Sparse (right column) targets. Shaded regions are rolling 95\% posterior predictive intervals and solid lines are posterior predictive medians. True observations are plotted as points. Observations exceeding the 97.5\% upper predictive boundary are highlighted to denote right-tail exceedances. The y-axis scales are unified within each column to facilitate direct comparisons of interval widths across models.}
\label{fig:sim_bands}
\smallskip
\alttext{Grid of time-series panels showing observed counts, median forecasts, and shaded 95\% predictive intervals for several models on dense and sparse simulated targets, with exceedances highlighted.}
\end{figure}

Aggregate error and tail calibration metrics (Table \ref{tab:sim_results}) summarize overall performance, but they can mask qualitatively different uncertainty behaviors that matter for monitoring. Figure \ref{fig:sim_bands} therefore plots the full rolling 95\% posterior predictive intervals, together with the median forecast and the realized counts, over the 50-step held-out window. Red markers indicate right-tail exceedances of the predictive 97.5\% quantile, which correspond to the anomaly rule used throughout this paper.

Several patterns in Figure \ref{fig:sim_bands} are consistent with, but not fully explained by, the aggregate metrics. First, the oracle (True DGP) provides a realistic reference for both sharpness and exceedances in a finite test window. Even under the correct data-generating mechanism, exceedances are random and integer-valued, so a nonzero exceedance count is expected in some runs. Visually, the oracle bands are neither uniformly tight nor uniformly wide. Instead, they expand during periods of genuine volatility and contract when the process is stable.

Second, the Full GLM illustrates a common failure mode in high-dimensional count regression without effective screening. Its predictive bands become inflated, especially for the sparse target, which reduces exceedances but produces thresholds that are operationally uninformative for anomaly detection. This is the main reason interval visualization is important alongside the tail calibration statistic $T$: a model can achieve a small $T$ via overly wide bands that rarely get exceeded, while still failing to separate routine fluctuations from genuinely unusual bursts.

Third, the AR2 baseline shows the opposite behavior. Because it only conditions on the target history, its predictive intervals are relatively smooth and can remain narrow during periods when the true process is being driven by cross-series spillovers. In the dense target, this appears as delayed adaptation around rapid level changes. In the sparse target, the AR2 intervals can stay close to zero and then miss sudden burst episodes, which manifests as exceedances that correspond to missed anticipatory uncertainty rather than true model overconfidence about the mean alone.

Fourth, the deep-representation hybrids exhibit two distinct regimes. The TFT-Bayesian head typically adapts its uncertainty in a manner closer to the oracle, consistent with its ability to learn nonlinear temporal structure. However, when the learned embedding is projected through an exponential link in the likelihood head, the resulting uncertainty can become wider and more variable in some segments, particularly after abrupt changes, reflecting sensitivity of the mean scale under the log link. The TSMixer-Bayesian head, in contrast, can produce intervals that are too tight in both dense and sparse settings, yielding clusters of exceedances. This visual pattern is consistent with under-dispersed predictive distributions, which can lead to elevated false-alarm rates when the same threshold rule is used for monitoring.

Fifth, the ``wide and deep'' variant (TSMixer + Shrinkage) clarifies why simply stacking neural embeddings with a high-dimensional shrinkage block can be unstable. When expressive embeddings and raw exogenous candidates overlap in the variation they can explain, collinearity can produce irregular interval behavior, including occasional over-expansion in the dense target without a commensurate improvement in exceedance control. This instability is visible as bands that widen sharply in short bursts rather than tracking sustained changes in conditional uncertainty.

Finally, the Two-Step Shrinkage GLM most closely reproduces the oracle-style adaptation of predictive uncertainty while remaining comparatively sharp. For the sparse target, its upper boundary stays close to zero during inactive periods and expands selectively during bursts, which is the desirable behavior for monitoring because it limits false positives during long flatlines while retaining coverage when the process becomes active. For the dense target, the Two-Step procedure yields bands that widen around transient peaks and contract afterward, reflecting the retention of a small set of true cross-series drivers while filtering out noise predictors that would otherwise inflate uncertainty.

Overall, Figure \ref{fig:sim_bands} complements Table \ref{tab:sim_results} by separating two practically distinct error modes: overly wide predictive intervals that obscure anomaly thresholds, and overly narrow intervals that generate excess exceedances. For the anomaly-detection use case motivating this work, achieving both reasonable tail frequency and informative sharpness is essential.

% =======================================================
\section{Case study: GDELT panels}
\label{sec:gdelt}
% =======================================================

We evaluate the proposed forecasting and monitoring pipelines on the two weekly GDELT panels introduced in \Cref{sec:data}. The empirical goal is twofold: (i) assess out-of-sample predictive accuracy for sparse and overdispersed event counts using likelihood-based probabilistic forecasts, and (ii) assess whether the fitted predictive distributions provide reliable right-tail behaviour for monitoring, since the intended operational use is to flag unusually large event counts for analyst triage.

Each target series is assigned an NB2 or ZINB2 likelihood head according to the training-period sparsity rule described in \Cref{sec:bayes_glm}. We report results separately for the resulting dense (NB2) and sparse (ZINB2) subsets, because these regimes pose different challenges. Dense series stress short-horizon tracking under overdispersion, whereas sparse series stress correct modelling of excess zeros and the conditional right tail when nonzero weeks occur.

\subsection{Predictive accuracy and tail calibration}
We compare the Two-Step Shrinkage GLM to two representation-learning baselines (TFT-Bayesian and TSMixer-Bayesian) using the test periods defined in \Cref{sec:data}. For the Israel-Palestine panel, the training period ends in January 2021, placing the October 2023 escalation inside the held-out test window. For the Russia-Ukraine panel, the training period ends in January 2020, placing the 2022 invasion period inside the held-out test window. This design ensures that the principal episodes discussed below are not used for model fitting or model selection.

Performance is summarised by the logarithmic mean absolute error (Log MAE) and the right-tail calibration metric $T$ (both defined in \Cref{sec:metrics}). Log MAE reflects median forecast accuracy on a stabilised scale and is less dominated by a small number of extreme weeks. The calibration metric $T$ targets monitoring behaviour by comparing the nominal right-tail exceedance probability (0.025 at $q=0.975$) to the empirical exceedance frequency on held-out data.

To ensure a stable like-for-like comparison, macro-averages are computed over the intersection of series for which all pipelines successfully produced posterior predictive distributions over the evaluation window. Table \ref{tab:metrics_gdelt_panels} reports mean and standard deviation of per-series metrics for both panels, separated into dense (NB2) and sparse (ZINB2) targets.

Two patterns are particularly relevant for monitoring. First, for dense targets, the Two-Step Shrinkage GLM yields the lowest Log MAE in both panels while maintaining competitive right-tail calibration. This behaviour is consistent with the model structure: the autoregressive block captures short-range dependence, while screening in the cross-series block reduces the influence of weak predictors that can destabilise the exponential mean mapping. Second, for sparse targets, TSMixer-Bayesian yields the lowest Log MAE on average, suggesting that shared multivariate representations can be effective for long stretches of zero activity. However, tail calibration remains essential for alerting: in both panels, the Two-Step Shrinkage GLM exhibits consistently strong calibration on the sparse subset, indicating reliable dynamic upper bounds for anomaly flagging without systematic over- or under-coverage of the right tail.

\subsection{Detected anomalies in a representative series}
\begin{figure}[tbp]
    \centering
    \includegraphics[width=0.85\linewidth]{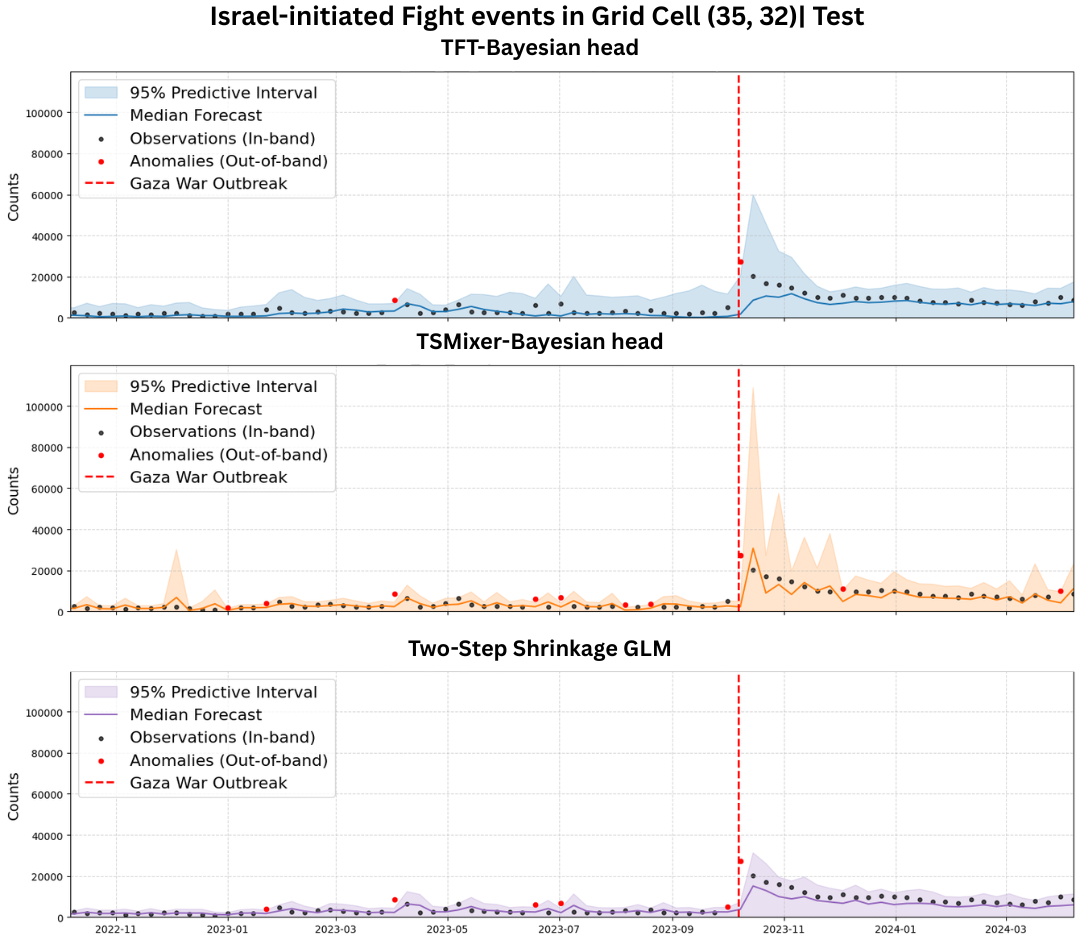}
    \caption{One-step-ahead posterior predictive summaries for Israel-initiated Fight events (CAMEO 19) in the target grid cell during the test period. The solid line is the posterior predictive median and the shaded band is the 95\% posterior predictive interval. Points are observed weekly event counts. Points above the 97.5\% predictive upper bound are highlighted. The vertical reference line marks October 2023.}
    \label{fig:isr_anomalies}
    
   \end{figure}

We illustrate how probabilistic outputs translate into monitoring decisions using a representative dense conflict series. Figure \ref{fig:isr_anomalies} shows one-step-ahead posterior predictive intervals and flagged right-tail exceedances for Fight events (CAMEO 19) initiated by Israel in the target region(35, 32) during the test period. The vertical marker highlights the October 2023 escalation, which constitutes a sharp regime change relative to pre-escalation activity. For the corresponding in-sample fit during training, see Figure \ref{fig:isr_anomalies_train}.

All pipelines produce early warnings in the weeks immediately preceding the escalation, with observed counts exceeding dynamic 97.5\% predictive upper bounds. After the regime change, the models differ in how quickly and how stably they adapt their predictive distributions. The Two-Step Shrinkage GLM exhibits a comparatively sharp pre-event envelope and then expands and recentres the predictive band to the elevated post-event level, maintaining informative intervals rather than permanently inflating uncertainty. The representation-learning pipelines, by contrast, tend to widen predictive envelopes more substantially after the break, which reduces alert specificity because broader bands imply fewer exceedances even when realised counts remain unusually high relative to pre-break behaviour.

\subsection{Directional spillover summaries for Israeli conflict series}
We next illustrate the interpretability benefits of the Two-Step Shrinkage GLM for directional spatial analysis. We focus on two dense target series in the Israel-Palestine panel: Assault events (CAMEO 18) and Fight events (CAMEO 19) initiated by Israel within the target region (the grid cell centred at longitude 35 and latitude 32). The screened cross-series component of the Step 1 model identifies a sparse set of lag-1 source series, and Step 2 refits the model on this active set to obtain posterior summaries for effects that remain credibly nonzero.

For directional interpretation, we restrict attention to active cross-series coefficients whose 95\% posterior credible intervals exclude zero (Table \ref{tab:spatial_ci}). Each selected source series is associated with a geographic source cell, so we map its coefficient to a geodesic bearing from the source centroid to the target centroid on the WGS84 ellipsoid (bearing computation described in \Cref{sec:directional_glm}). This produces two complementary directional summaries.

\begin{figure}[htbp]
    \centering
    \includegraphics[width=0.9\linewidth]{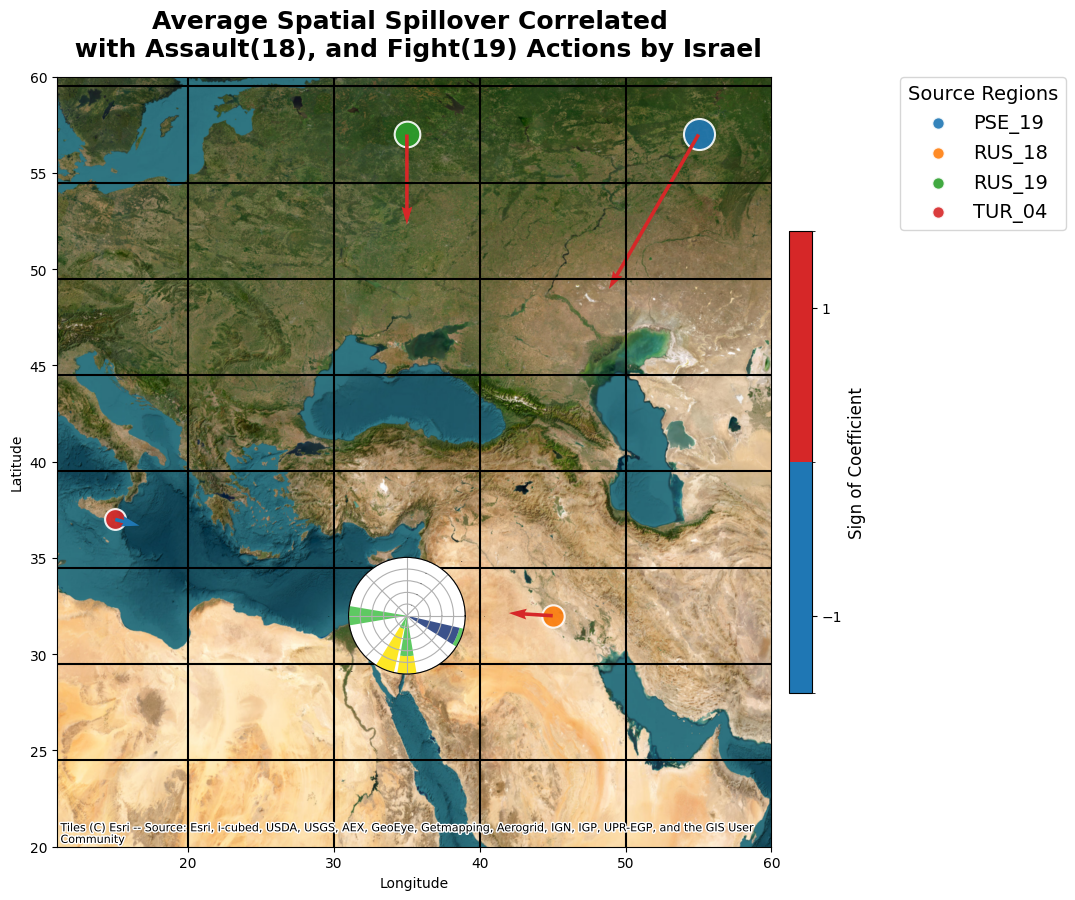}
    \caption{Geographic map of selected lag-1 spillover sources for Israeli Assault (CAMEO 18) and Fight (CAMEO 19) targets in the Israel-Palestine panel. Points mark source cell centroids. Point size is proportional to the absolute value of the posterior mean cross-series coefficient. Arrows indicate geodesic bearings from each source centroid to the target centroid. Arrow colour indicates the sign of the posterior mean effect.}
    \label{fig:spatial_map}
    \smallskip
    \alttext{Geographic map showing a target grid cell and multiple source cells; arrows point from sources to target and indicate bearing; marker size indicates coefficient magnitude and marker styling indicates sign.}
\end{figure}

Figure \ref{fig:spatial_map} displays the selected sources on the geographic domain. The map shows where the active sources are located relative to the target cell, and it encodes both the direction of each source (via the geodesic arrow from source to target) and the estimated effect size and sign (via the point size and arrow colour). This view supports event-level diagnosis, since specific sources can be traced to specific regions and event types listed in Table \ref{tab:spatial_ci}.

\begin{figure}[htbp]
    \centering
    \begin{minipage}[t]{0.48\linewidth}
        \centering
        \includegraphics[width=\linewidth]{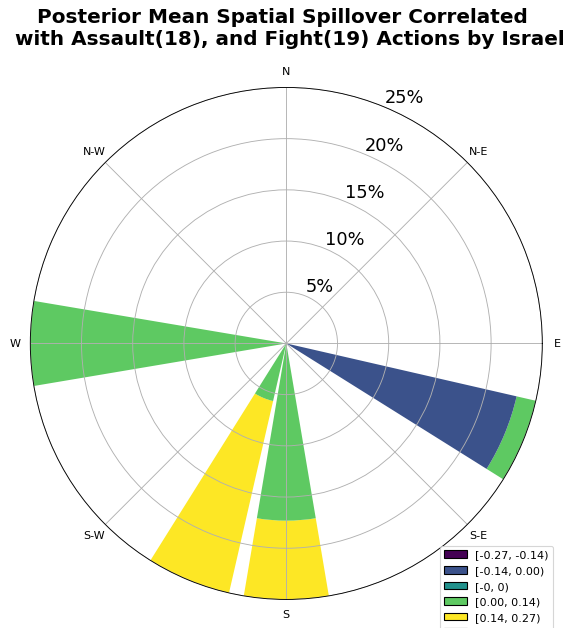}
        \caption{Windrose summary of the selected spillover sources. This plot provides an aggregated directional view of the spatial effects shown in Figure \ref{fig:spatial_map}. Petal direction denotes the bearing from source to target, and petal length is proportional to the number of selected sources within each bearing sector. Shading encodes binned posterior mean coefficient magnitudes.}
        \label{fig:spatial_windrose}
        \smallskip
        \alttext{Circular windrose chart summarising the bearings of selected spillover sources toward the target, with petal lengths reflecting counts per bearing sector and shading indicating binned coefficient magnitudes.}
    \end{minipage}%
    \hfill
    \begin{minipage}[t]{0.48\linewidth}
        \centering
        \includegraphics[width=\linewidth]{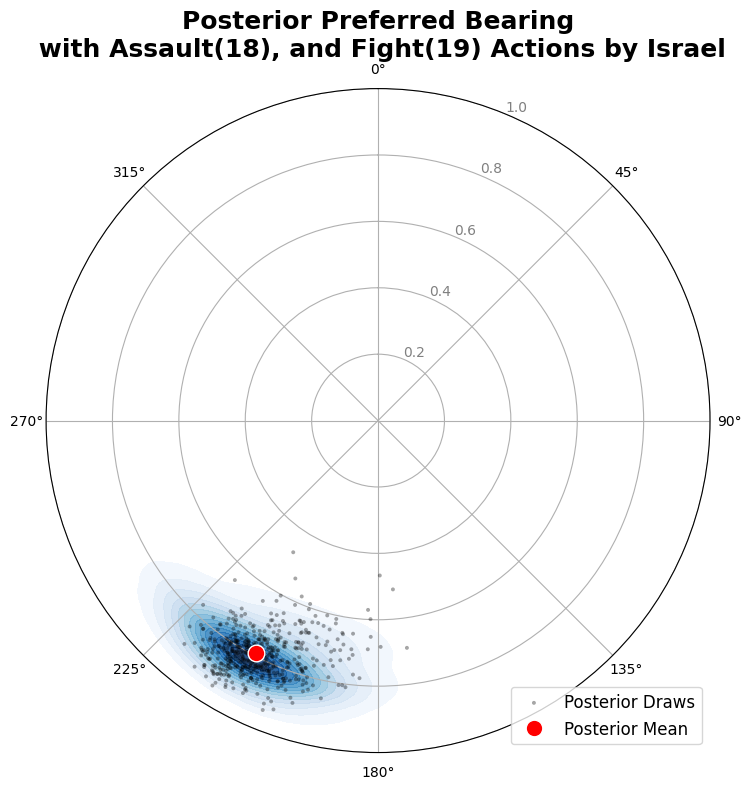}
        \caption{Posterior density of the preferred bearing ($\Omega_i$) and directional concentration ($R_i$). The blue contours represent a 2D kernel density estimate of the posterior draws, illustrating the model's high certainty regarding the spillover direction. The red dot indicates the posterior mean.}
        \label{fig:polar_kde}
        \smallskip
        \alttext{Polar contour plot showing a concentrated blue probability density near the southwest edge, indicating a high certainty and high concentration of the preferred bearing from the posterior draws.}
    \end{minipage}
\end{figure}

Figure \ref{fig:spatial_windrose} summarises the same set of selected sources on the circle by aggregating them into bearing sectors. The windrose emphasises directional concentration: it shows whether selected spillovers are clustered into a few preferred directions or distributed broadly around the compass, and it provides a compact view of the distribution of effect magnitudes by direction.

To explicitly quantify the uncertainty of this aggregate directional effect, we visualize the joint posterior distribution of the preferred bearing $\Omega_i^{(s)}$ and the directional concentration $R_i^{(s)}$, as defined in \Cref{sec:directional_glm}. Figure \ref{fig:polar_kde} projects the posterior draws onto a polar coordinate system using a 2D kernel density estimate. The dense probability mass (blue contours) near the boundary confirms that the model is highly certain of the estimated spillover direction, while the radial distance approaching 1.0 indicates a strong directional consensus among the active sources.

These summaries are conditional associations under the fitted regression model rather than causal effects. They identify cross-series signals that improve short-horizon forecasting while remaining geographically interpretable through bearings, and they provide context for flagged monitoring windows by indicating which directions and neighbouring sources are most strongly associated with the target series under the fitted model.

\section{Discussion and conclusion}
\label{sec:disc}

This paper develops a modular framework for forecasting and monitoring sparse, bursty event-count panels derived from the Global Database of Events, Language and Tone. The first contribution is a likelihood-based approach to uncertainty: deep temporal encoders are paired with negative binomial or zero-inflated negative binomial likelihood heads so that predictive intervals and exceedance probabilities are obtained from a fully specified count model rather than from ad hoc uncertainty surrogates. This design supports rare-event monitoring because it yields posterior predictive right-tail probabilities that are directly interpretable as alert scores.

A second contribution is interpretable spillover attribution in high dimensions. The Bayesian generalised linear model uses lagged cross-series predictors with three-parameter beta-normal shrinkage and a two-step screen-and-refit procedure. Screening provides an interpretable active set of spillover drivers, while refitting reduces shrinkage bias on retained effects and yields a sparse posterior that can be used for forecasting and monitoring.

A third contribution is a directional layer for interpretation on the sphere. Active cross-region effects from the sparse spillover model are mapped to geodesic bearings on the World Geodetic System 1984 ellipsoid. Weighted circular summaries and visualisations, including bearing-field maps and rose diagrams, provide compact diagnostics that link anomaly flags to geographically interpretable directions and source regions.

Across simulation and empirical panels, the results emphasise tail calibration as a primary diagnostic for monitoring. Point forecast accuracy is important, but operational alerting depends on whether predictive upper bounds are exceeded at the intended frequency. The likelihood-head construction supports this diagnostic directly by producing predictive quantiles and exceedance probabilities under a fitted count distribution, and the two-step spillover model provides a practical compromise between predictive performance and interpretability.

Several extensions are natural. Spatial neighbourhood construction could be made adaptive, for example by allowing distance-dependent candidate sets that vary by target region. The robustness of anomaly scoring to reporting intensity shifts and to changes in media coverage warrants further study. Finally, joint modelling that propagates uncertainty from the encoder stage into the likelihood head could further improve probabilistic calibration in settings where representation error is substantial.

Overall, the proposed pipelines provide a coherent workflow for calibrated forecasting, anomaly scoring, and directional spillover interpretation in high-dimensional spatiotemporal event-count panels.

% =========================
% Tables (end-of-text block)
% =========================

% Table first cited in \Cref{sec:sim_results}
\begin{table}[p]
    \centering
    \caption{Predictive performance in the simulation study on the 50-week test window. MAE (Raw) is on the weekly event-count scale; MAE (Log) is on the $\log_{10}(1+y)$ scale; Tail calibration $T$ is computed at the 97.5\% predictive upper bound.}
    \label{tab:sim_results}
    \begin{tabular}{ll
                    S[table-format=1.2]
                    S[table-format=1.4]
                    S[table-format=1.3]}
        \toprule
        {Target} & {Model} & {MAE (Raw)} & {MAE (Log)} & {Tail calibration $T$} \\
        \midrule
        Dense  & True DGP                 & 2.50 & 0.2741 & 0.035 \\
        Dense  & AR2 Bayesian baseline     & 2.62 & 0.2904 & 0.005 \\
        Dense  & Full Bayesian GLM         & 2.53 & 0.2725 & 0.005 \\
        Dense  & TFT-Bayesian              & 2.44 & 0.2686 & 0.035 \\
        Dense  & TSMixer-Bayesian          & 3.26 & 0.3505 & 0.175 \\
        Dense  & Two-Step Shrinkage        & 2.54 & 0.2739 & 0.035 \\
        Dense  & TSMixer plus shrinkage    & 3.16 & 0.3413 & 0.155 \\
        \addlinespace[6pt]
        Sparse & True DGP                 & 0.76 & 0.1498 & 0.025 \\
        Sparse & AR2 baseline              & 0.78 & 0.1604 & 0.025 \\
        Sparse & Full Bayesian GLM         & 1.08 & 0.2102 & 0.005 \\
        Sparse & TFT-Bayesian              & 0.76 & 0.1544 & 0.025 \\
        Sparse & TSMixer-Bayesian          & 0.86 & 0.1784 & 0.095 \\
        Sparse & Two-Step Shrinkage        & 0.86 & 0.1739 & 0.025 \\
        Sparse & TSMixer plus shrinkage    & 0.84 & 0.1734 & 0.055 \\
        \bottomrule
    \end{tabular}
\end{table}

% Table first cited in \Cref{sec:gdelt}
\begin{table}[p]
\centering
\caption{Macro-averaged predictive performance on the GDELT test sets. Results are reported as mean (standard deviation) across series. Lower values indicate better performance for all metrics.}
\label{tab:metrics_gdelt_panels}
\begin{tabular}{l r r r r}
\toprule
& \multicolumn{2}{c}{\textbf{Dense targets (NB2)}} & \multicolumn{2}{c}{\textbf{Sparse targets (ZINB2)}} \\
\cmidrule(lr){2-3} \cmidrule(lr){4-5}
\textbf{Panel and model} & \textbf{MAE (log)} & \textbf{Tail calibration $T$} & \textbf{MAE (log)} & \textbf{Tail calibration $T$} \\
\midrule
\multicolumn{5}{l}{\textbf{Israel-Palestine panel (971 series)}} \\
\addlinespace[2pt]
TFT-Bayesian          & 0.411 (0.123) & 0.065 (0.090) & 0.263 (0.807) & 0.048 (0.079) \\
Two-Step Shrinkage    & \textbf{0.392} (0.192) & 0.052 (0.096) & 0.354 (0.921) & 0.044 (0.080) \\
TSMixer-Bayesian      & 0.522 (0.164) & \textbf{0.042} (0.048) & \textbf{0.241} (0.266) & \textbf{0.042} (0.064) \\
\addlinespace[6pt]
\multicolumn{5}{l}{\textbf{Russia-Ukraine panel (969 series)}} \\
\addlinespace[2pt]
TFT-Bayesian          & 0.584 (0.324) & 0.085 (0.130) & 0.234 (0.611) & 0.050 (0.089) \\
Two-Step Shrinkage    & \textbf{0.391} (0.206) & \textbf{0.043} (0.078) & 0.275 (0.755) & \textbf{0.040} (0.072) \\
TSMixer-Bayesian      & 0.883 (0.429) & 0.088 (0.101) & \textbf{0.230} (0.280) & 0.053 (0.077) \\
\bottomrule
\end{tabular}
\end{table}

% Table first cited in directional spillover subsection of \Cref{sec:gdelt}
\begin{table}[p]
\centering
\caption{Selected cross-series effects for Israeli conflict targets. Entries list sources whose 95\% posterior credible intervals exclude zero under the Step 2 refitted Two-Step Shrinkage GLM.}
\label{tab:spatial_ci}
\begin{tabular}{llll}
\toprule
\textbf{Target series} & \textbf{Source region} & \textbf{Event type} & \textbf{95\% CI} \\
\midrule
Assault (CAMEO 18) & Palestine (55,57) & Fight (19)   & $[0.0274,\ 0.4968]$ \\
Assault (CAMEO 18) & Russia (45,32)    & Assault (18) & $[0.0048,\ 0.0948]$ \\
Assault (CAMEO 18) & Calendar effect   & Week 21      & $[-0.5797,\ -0.0110]$ \\
\addlinespace[4pt]
Fight (CAMEO 19)   & Russia (35,57)    & Fight (19)   & $[0.0319,\ 0.2025]$ \\
Fight (CAMEO 19)   & Turkey (15,37)    & Consult (04) & $[-0.0691,\ -0.0018]$ \\
\bottomrule
\end{tabular}
\end{table}

\begin{appendices}

\section{Computation and posterior inference details}
\label{app:posterior_inference}

This appendix summarizes implementation choices used in our \texttt{NumPyro} models, focusing on (i) stable parameterizations for NB2 and ZINB2 likelihoods, (ii) a non-centered implementation of TPBN shrinkage that is well behaved under NUTS, and (iii) posterior predictive simulation used for forecast quantiles and tail probabilities.

\subsection{Likelihood parameterizations and numerical stability}
\label{app:likelihood_mapping}

\textbf{NB2 mean-dispersion form.}
Throughout, NB2 is parameterized by a mean $\mu>0$ and a dispersion $\alpha>0$, so that
\[
\mathrm{Var}(Y\mid \mu,\alpha)=\mu+\alpha\mu^2.
\]
In computation we work with the equivalent concentration parameter
\[
\kappa=\frac{1}{\alpha+\varepsilon},\qquad \varepsilon=10^{-5},
\]
to avoid numerical issues when $\alpha$ is close to $0$.

\textbf{ZINB2 mixture likelihood in log space.}
For ZINB2, we combine a Bernoulli structural-zero gate with the NB2 count component:
\[
Y \sim
\begin{cases}
0, & \text{with probability }\pi,\\
\mathrm{NB2}(\mu,\alpha), & \text{with probability }1-\pi.
\end{cases}
\]
We compute the log-likelihood using a stable log-sum-exp expression at $Y=0$.
Let $\pi$ be clipped to $[\varepsilon,1-\varepsilon]$. Writing $p_{\mathrm{NB2}}(\cdot\mid\mu,\alpha)$ for the NB2 pmf, the per-observation log-likelihood is
\begin{align}
\log p(Y\mid \pi,\mu,\alpha)
&=
\mathbb{I}\{Y=0\}\,
\operatorname{logsumexp}\!\Big(\log\pi,\; \log(1-\pi)+\log p_{\mathrm{NB2}}(0\mid\mu,\alpha)\Big)
\nonumber\\
&\quad+
\mathbb{I}\{Y>0\}\,\Big(\log(1-\pi)+\log p_{\mathrm{NB2}}(Y\mid\mu,\alpha)\Big),
\end{align}
where $\operatorname{logsumexp}(a,b)=\log(\exp(a)+\exp(b))$.

\textbf{Link functions and clipping.}
For both NB2 and ZINB2, the mean predictor is $\mu=\exp(\eta)$ with $\eta=X\beta+Z\gamma$.
Before exponentiation we clip
\[
\eta \leftarrow \min\big(\max(\eta,\eta_{\min}),\eta_{\max}\big),
\qquad \eta_{\min}=-12,\ \eta_{\max}=10,
\]
to avoid overflow and to stabilize gradients. For the ZINB2 gate we use $\pi=\operatorname{logit}^{-1}(\eta_\pi)$ with analogous clipping of $\eta_\pi$ if needed, and with $\pi$ additionally clipped to $[\varepsilon,1-\varepsilon]$ as above.

\subsection{TPBN shrinkage prior: non-centered implementation}
\label{app:tpbn_reparam}

The TPBN family can be expressed as a global-local scale mixture, but naive centered parameterizations can exhibit funnel-shaped geometry that causes divergent transitions under NUTS. We therefore use a non-centered form based on an auxiliary Beta draw.

Let $\gamma_j$ denote a shrinkage coefficient. We introduce
\[
\xi_j \sim \mathrm{Beta}(u,a),
\qquad
\lambda_j^2=\frac{1-\xi_j}{\xi_j+\varepsilon},
\qquad
\tau \sim \mathrm{HalfCauchy}(0,\tau_0),
\]
and sample coefficients as
\[
z_j\sim \mathcal{N}(0,1),
\qquad
\gamma_j=\tau\,\lambda_j\,z_j.
\]
This parameterization retains the intended heavy-tailed global-local shrinkage behavior while improving posterior geometry for HMC.

\subsection{Posterior predictive simulation}
\label{app:posterior_predictive}

Forecast quantiles and tail probabilities are computed from posterior predictive draws. For each posterior draw $\Theta^{(s)}$ and each forecasted time point, we compute $(\mu^{(s)},\alpha^{(s)})$ (and $\pi^{(s)}$ under ZINB2), then simulate:
\begin{align}
\tilde{Y}^{(s)}_{\mathrm{NB2}} &\sim \mathrm{NB2}\!\left(\mu^{(s)},\alpha^{(s)}\right),
\\
\tilde{Y}^{(s)}_{\mathrm{ZINB2}}
&=
(1-Z^{(s)})\,\tilde{Y}^{(s)}_{\mathrm{NB2}},
\qquad
Z^{(s)}\sim \mathrm{Bernoulli}\!\left(\pi^{(s)}\right).
\end{align}
Given draws $\{\tilde{Y}^{(s)}\}_{s=1}^S$, predictive quantiles (for example $U_{t,0.975}$) are computed as empirical quantiles, and predictive right-tail probabilities are computed as Monte Carlo averages of $\mathbb{I}\{\tilde{Y}^{(s)}\ge Y_{\mathrm{obs}}\}$.

\section{Algorithms for fitted pipelines}
\label{app:algorithms}

Algorithm \ref{alg:twostep_nuts} summarizes the Two-Step screen-and-refit procedure used for the TPBN-regularized GLM. Algorithms \ref{alg:tsm_hybrid} and \ref{alg:tsm_twostep_nuts} summarize the encoder-plus-likelihood-head variants used in benchmarking.

\begin{algorithm}[tbp]
\caption{Two-Step Bayesian Shrinkage Inference (NB2 or ZINB2)}
\label{alg:twostep_nuts}
\begin{algorithmic}[1]
\Require Response vector $Y$, fixed-effects matrix $H_{\mathrm{fixed}}$, candidate spillover matrix $H_{\mathrm{shrink}}$, global scale $\tau_0$, selection tolerance $\delta\ge 0$.
\Statex \textbf{Phase 1: Shrinkage fit (screening)}
\State Fit NB2 or ZINB2 with linear predictors
\[
\eta = H_{\mathrm{fixed}}\beta + H_{\mathrm{shrink}}\bm{\gamma},
\qquad
\eta_\pi = H_{\mathrm{fixed}}\beta^{(\pi)} + H_{\mathrm{shrink}}\bm{\gamma}^{(\pi)} \ \text{(ZINB2 only)},
\]
with Gaussian priors on $\beta$ (and $\beta^{(\pi)}$) and TPBN priors on $\bm{\gamma}$ (and $\bm{\gamma}^{(\pi)}$), using NUTS.
\State Compute marginal $95\%$ credible intervals $[L_j,U_j]$ for each $\gamma_j$ (and each $\gamma^{(\pi)}_j$ if ZINB2).
\State Define the active set
\[
\mathcal{A}
=
\left\{j:\; L_j>\delta \ \text{or}\ U_j<-\delta\right\}
\ \cup\
\left\{j:\; L^{(\pi)}_j>\delta \ \text{or}\ U^{(\pi)}_j<-\delta\right\}
\ \text{(second term for ZINB2 only)}.
\]
\Statex \textbf{Phase 2: Refit without shrinkage (estimation)}
\If{$|\mathcal{A}|=0$}
\State Set $H_{\mathrm{step2}}=H_{\mathrm{fixed}}$.
\Else
\State Set $H_{\mathrm{step2}}=[\,H_{\mathrm{fixed}},\ H_{\mathrm{shrink}}[:,\mathcal{A}]\,]$.
\EndIf
\State Refit NB2 or ZINB2 using $H_{\mathrm{step2}}$ with Gaussian priors (no TPBN), using NUTS.
\State Generate posterior predictive draws $\tilde{Y}$ for forecasting and anomaly scoring.
\State \Return Posterior samples and posterior predictive draws.
\end{algorithmic}
\end{algorithm}

\begin{algorithm}[tbp]
\caption{TSMixer-Hybrid Bayesian Inference (encoder plus likelihood head)}
\label{alg:tsm_hybrid}
\begin{algorithmic}[1]
\Require Target series $Y$, multivariate history tensor $X_{\mathrm{hist}}$, AR design $H_{\mathrm{AR}}$, trained TSMixer encoder $g_{\mathrm{TSM}}$.
\State Freeze the trained encoder parameters and compute target embeddings $E = g_{\mathrm{TSM}}(X_{\mathrm{hist}})$.
\State Form the likelihood-head design matrix $H_{\mathrm{fixed}}=[\,H_{\mathrm{AR}},\ E\,]$.
\State Fit the Bayesian NB2 or ZINB2 likelihood head via NUTS with Gaussian priors on head coefficients.
\State Simulate posterior predictive draws $\tilde{Y}$ for quantiles and tail probabilities.
\State \Return Posterior samples and posterior predictive draws.
\end{algorithmic}
\end{algorithm}

\begin{algorithm}[tbp]
\caption{Two-Step Bayesian Inference (TSMixer + Shrinkage)}
\label{alg:tsm_twostep_nuts}
\begin{algorithmic}[1]
\Require Response $Y$, history tensor $X_{\mathrm{hist}}$, AR design $H_{\mathrm{AR}}$, candidate spillovers $H_{\mathrm{shrink}}$, trained encoder $g_{\theta}$, global scale $\tau_0$, selection tolerance $\delta$.
\State Compute embeddings $E=g_{\theta}(X_{\mathrm{hist}})$ and set $H_{\mathrm{fixed}}=[\,H_{\mathrm{AR}},\ E\,]$.
\State Apply Algorithm \ref{alg:twostep_nuts} with this $H_{\mathrm{fixed}}$ and $H_{\mathrm{shrink}}$ (TPBN screening then refit without shrinkage).
\State \Return Posterior samples and posterior predictive draws.
\end{algorithmic}
\end{algorithm}

\begin{figure}[tbp]
    \centering
    \includegraphics[width=0.75\linewidth]{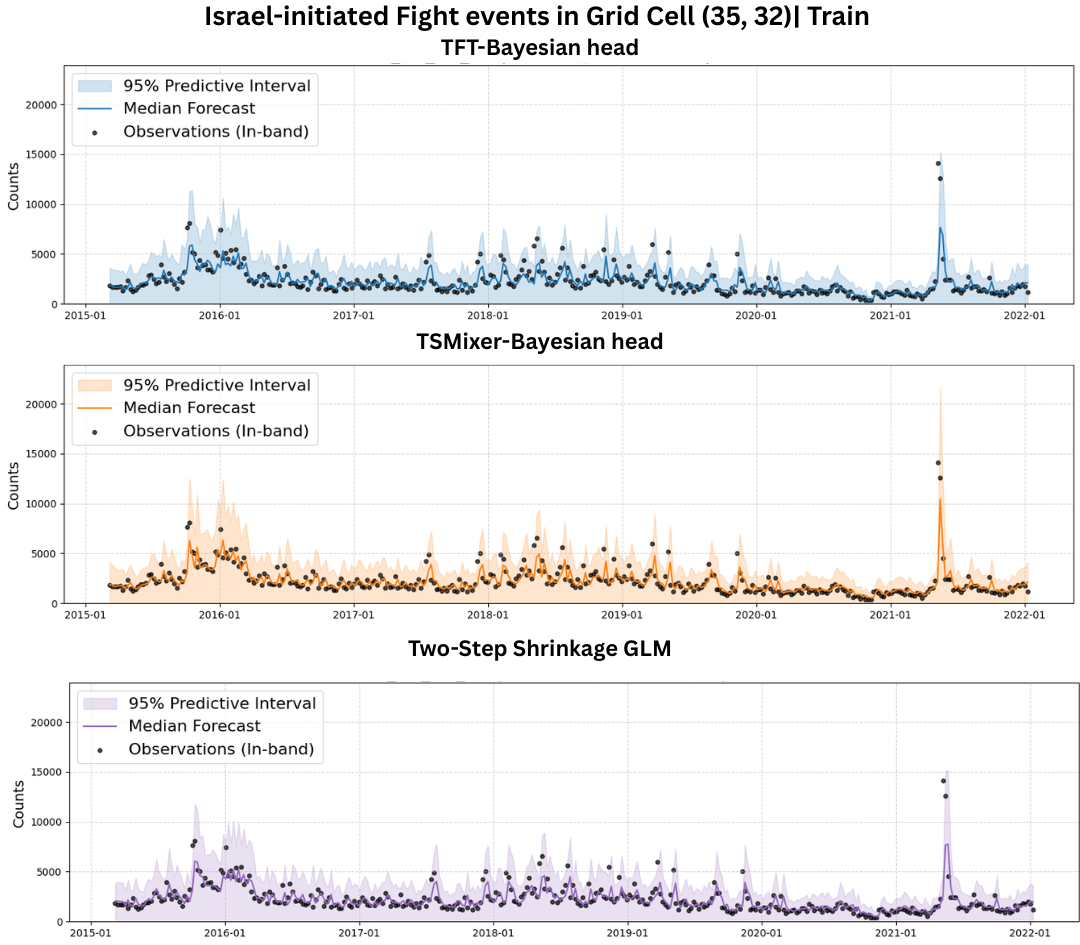}
    \caption{In-sample fit during the training period for the Israel Fight series used in the case study. The plot shows the fitted median trajectory and the associated posterior predictive interval over the training window.}
    \label{fig:isr_anomalies_train}
    \smallskip
\alttext{Time-series plot over the training period showing observed counts with a fitted median trajectory and shaded predictive interval.}
\end{figure}

\section{Hyperparameters and experimental constants}
\label{app:hyperparameters}

\subsection{Bayesian inference and prior specification}
\label{app:hyperparameters_bayes}
All Bayesian generalized linear models and likelihood heads were implemented using \texttt{NumPyro} and fitted via the No-U-Turn Sampler (NUTS). 
\begin{itemize}
    \item \textbf{MCMC settings:} For the simulation study, NUTS was run using 1 chain with 1,000 warmup steps and 6,000 posterior samples. For the empirical GDELT panels, the models utilized 500 warmup steps and 10,000 posterior samples to ensure stable convergence across the high-dimensional spatial predictors.
    \item \textbf{Prior distributions:} The unpenalized fixed effects ($\beta, \beta'$) were assigned weakly informative Gaussian priors, $\mathcal{N}(0, \sigma^2)$ with $\sigma = 100.0$. The negative binomial dispersion parameter $\alpha$ was assigned a $\mathrm{Gamma}(1.0, 10.0)$ prior. 
    \item \textbf{TPBN shrinkage:} The hierarchical Gamma-Gamma-Normal shrinkage parameters were set to $u = 0.5$, $a = 0.5$, and a global scale of $\tau_0 = 0.5$. 
\end{itemize}

\subsection{Data generating process (DGP) constants for simulation}
\label{app:hyperparameters_dgp}
The ground-truth synthetic data was generated over $T=1000$ steps for 100 parallel series (1 dense target, 1 sparse target, and 98 Poisson noise covariates generated with $\lambda = 1.5$). The true dispersion parameter was set to $\alpha = 0.5$. To prevent numerical overflow during the exponential link projection in the DGP, the true linear predictors were clipped to $[-15.0, 15.0]$. The zero-inflation probabilities were clipped to $[10^{-6}, 1 - 10^{-6}]$. All lagged features were transformed using $\log(1+x)$ before entering the linear predictors. The precise active coefficients are as follows:

\begin{itemize}
    \item \textbf{Dense target (NB2):} The unpenalized autoregressive block (intercept, lag-1, lag-2) was set to $\beta_{\mathrm{fixed}}^{(D)} = [0.5, 0.2, 0.1]^{\top}$. Among the 99 cross-series shrinkage candidates, the 5 active signals (lag-1 of the sparse target and lag-1 of the first 4 noise covariates) were assigned coefficients: $\gamma_{\mathrm{active}}^{(D)} = [0.4, 0.6, -0.5, 0.5, -0.6]^{\top}$.
    \item \textbf{Sparse target (ZINB2):} 
    \begin{itemize}
        \item \textit{Count component ($\mu$):} The unpenalized autoregressive block was set to $\beta_{\mathrm{fixed}}^{(S)} = [0.2, 0.1, 0.05]^{\top}$. The corresponding 5 active cross-series covariates were assigned coefficients: $\gamma_{\mathrm{active}}^{(S)} = [0.5, -0.6, 0.4, -0.5, 0.6]^{\top}$.
        \item \textit{Zero-inflation gate ($\pi$):} The autoregressive block was set to $\gamma_{\mathrm{fixed}}^{(S)} = [-1.0, 0.2, 0.1]^{\top}$. The corresponding 5 active cross-series covariates were assigned coefficients: $\gamma_{\mathrm{active}}^{\prime (S)} = [0.6, 0.7, -0.5, 0.6, -0.7]^{\top}$.
    \end{itemize}
\end{itemize}

\subsection{Deep temporal encoders}
\label{app:hyperparameters_encoders}
The representation learning baselines utilized frozen embeddings extracted from pre-trained Temporal Fusion Transformer (TFT) and Time Series Mixer (TSMixer) networks. 

\begin{itemize}
    \item \textbf{Architecture specifications:} 
    \begin{itemize}
        \item \textit{TFT:} Across both simulation and GDELT panels, the TFT was configured with a state dimension of 24, 2 LSTM layers, 3 attention heads, and a dropout rate of 0.05.
        \item \textit{TSMixer:} For the simulation dataset, TSMixer utilized a lookback window of 10, 3 mixer blocks, and a feed-forward dimension of 24. For the Israel-Palestine (ISR) panel, the lookback was set to 5 with a feed-forward dimension of 19. For the Russia-Ukraine (UKR) panel, the lookback was expanded to 29 with a feed-forward dimension of 48.
    \end{itemize}
    \item \textbf{Training hyperparameters:} Both models were trained using the Adam optimizer. The learning rate for the TFT models was initialized at $4 \times 10^{-5}$, while the TSMixer utilized $5 \times 10^{-5}$. Both architectures utilized an ExponentialLR scheduler with a decay rate of $\gamma = 0.97$, stepping every 100 to 200 epochs.
    \item \textbf{Batch sizes and epochs:} During training, the TFT utilized a batch size of 64 for the simulation data and 32 for the empirical panels. The TSMixer utilized a batch size of 32 universally. The TFT was trained for up to 10,000 epochs in the simulation setting and 3,000 epochs for the GDELT panels. The TSMixer was trained for up to 8,000 epochs. All models utilized early stopping heuristics based on validation set performance under a quantile loss objective.
\end{itemize}

\end{appendices}

% =======================================================
% Back matter statements
% =======================================================
\section*{Data availability}
GDELT event records are publicly available. Code to download, preprocess, and aggregate the raw GDELT feeds into the weekly spatiotemporal panels used in this paper is available at \href{https://github.com/KingJMS1/GDELTAnomalyAnalysis}{https://github.com/KingJMS1/GDELTAnomalyAnalysis}. The specific aggregated panel dataset used in our experiments is released in the same repository.

\section*{Code availability}
A \texttt{PyTorch} implementation of the deep encoders and a \texttt{NumPyro} implementation of the Bayesian NB2/ZINB2 models (including TPBN shrinkage, posterior predictive simulation, and visualization scripts) are available at \href{https://github.com/KingJMS1/GDELTAnomalyAnalysis}{https://github.com/KingJMS1/GDELTAnomalyAnalysis}.

\section*{Competing interests}
The authors declare no competing interests.

\section*{Author contributions}
H.H.H. conceived the study. H.H.H., M.S., and Y.H.C. developed the methodology. Y.H.C. and M.S. trained the models and ran the experiments. All authors contributed to interpretation of results and writing.

\section*{Acknowledgments}
H.-H. Huang was supported in part by the United States National Science Foundation (NSF) grants DMS-1924792 and DMS-23189250.

\bibliographystyle{oup-abbrvnat}
\bibliography{references}

\end{document}